\documentclass{acmart}
\usepackage{graphicx}
\usepackage{graphicx}
\usepackage{subcaption}
\usepackage{comment}
\usepackage{caption}
\usepackage{siunitx}
\usepackage{booktabs}
\usepackage{pifont}
\usepackage{hyphenat}
\newcommand{\cmark}{\ding{51}}%
\newcommand{\xmark}{\ding{55}}
\graphicspath{{./images/}}
\usepackage{pdfpages}

\acmJournal{TCPS}

\begin{document}

\title{Experiences \& Challenges with Server-Side WiFi Indoor Localization Using Existing Infrastructure}

\author{Dheryta Jaisinghani}
\affiliation{%
  \institution{University of Northern Iowa}
  \city{Cedar Falls, IA}
  \country{USA}
}
\email{dheryta.jaisinghani@uni.edu}

\author{Vinayak Naik}
\affiliation{%
  \institution{BITS Pilani Goa}
  \city{Goa}
  \country{India}}
\email{vinayak@goa.bits-pilani.ac.in}

\author{Rajesh Balan}
\affiliation{%
  \institution{Singapore Management University}
  \city{Singapore}
  \country{Singapore}
}
\email{rajesh@smu.edu.sg}

\author{Archan Misra}
\affiliation{%
  \institution{Singapore Management University}
  \city{Singapore}
  \country{Singapore}
}
\email{archanm@smu.edu.sg}

\author{Youngki Lee}
\affiliation{%
  \institution{Seoul National University}
  \city{Seoul}
  \country{South Korea}
}
\email{youngkilee@snu.ac.kr}

\begin{abstract}

Real-world deployments of WiFi-based indoor localization in large public
venues are few and far between as most state-of-the-art solutions require
either client or infrastructure-side changes.  Hence, even though high
location accuracy is possible with these solutions, they are not practical
due to cost and/or client adoption reasons. Majority of the public venues
use commercial controller-managed WLAN solutions, 
that neither allow client changes nor infrastructure changes.
In fact, for such venues we have observed highly heterogeneous devices  with very low adoption rates for client-side apps.

In this paper, we present our experiences in deploying a scalable
location system for such venues. 
We show that server-side localization
is not trivial and present two unique challenges associated with this
approach, namely \emph{Cardinality Mismatch} and \emph{High Client Scan
Latency}.  The ``Mismatch'' challenge results in a significant mismatch
between the set of access points (APs) reporting a client in the offline and
online phases, while the ``Latency'' challenge results in a low number of
APs reporting data for any particular client.  
We collect three weeks of detailed ground truth data ($\approx 200$ landmarks), 
from a WiFi setup that has been deployed for more than four years,
to provide evidences for the extent and understanding the impact of these problems.
Our analysis of real-world client devices reveal that the current trend for the clients is to reduce scans, thereby adversely impacting their localization accuracy. We analyze how localization is impacted when scans are minimal.
We propose heuristics to alleviate reduction in the accuracy despite lesser scans. Besides the number of scans, we summarize the other challenges and
pitfalls of real deployments which hamper the localization accuracy.

\end{abstract}



\vspace{1em}

\maketitle

\section{Introduction}
\label{sec:introduction}

There has been a long and rich history of WiFi-based indoor localization research~\cite{centaur, will, zee, localizeWithoutInfra, virtualCompass, placeLab, surroundSense, unsupervisedIndoorLoc, pushTheLimit, horus, monalisa, minimizingcalibration, radar, wigem, localizeWithoutPreDeployment, apsUsingAoA, zeroStartupCosts, phaser, arraytrack, spotFi, pinpoint, sensorToA, caesar, sail, cupid, wifiImaging, humanActivity, reusing60GHz, seeThroughWalls, mtrack}.  However, in spite of several breakthroughs, there are very few real-world deployments of WiFi-based indoor localization systems in public spaces.  The reasons for this are many-fold, with three of the most common being -- ($a$) the high cost of deployment, ($b$) arguably, the lack of compelling business use, and ($c$) the inability of existing solutions to seamlessly work with all devices. 
In fact, current solutions impose a tradeoff between universality, accuracy, and energy, for example, client-based solutions that combine inertial-based tracking with WiFi scanning offer significantly better accuracy but require a mobile application which will possibly drain energy faster and which will be downloaded by only a fraction of visitors~\cite{localizationSurvey}.

In this paper, we present our experiences with deploying and operating a WiFi-based indoor localization system across the entire campus of a small Asian university. 
It is worth noting that the environment is very densely occupied, by $\approx$  $10,000$ students and $1,500$ faculty and staff.
The system has been in the production for more than four years.
It is deployed at multiple venues including two universities (Singapore Management University, University of Massachusetts, Amherst), and four different public spaces (Mall, Convention Center, Airport, and Sentosa Resort)~\cite{livelabs_mobisys, livelabs_hotmobile}.
These venues use the localization system for various real-time analytics such as group detection, occupancy detection, and queue detection while taking care of user privacy.

Our goal is to highlight challenges and propose easy to integrate solutions to build a universal indoor localization system -- one that can spot localize all WiFi enabled devices on campus without any modifications whether client or infrastructure-side.
The scale and the nature of this real environment, presents unique set of challenges -- ($a$) infrastructure \textit{i.e.} controller and APs do not allow any changes, ($b$) devices cannot be modified in any way \textit{i.e.} no explicit/implicit participation for data generation, no app download allowed, and no chipset changes allowed, and ($c$) only available data is RSSI measurements from APs, which are centrally controlled by the controller, using a \emph{Real-Time Location Service} (RTLS) interface~\cite{rtls}.
It is worth noting that within the face of these challenges 
we have to rule out more sophisticated state-of-the-art schemes, such as fine-grained CSI measurements~\cite{spotFi}, Angle-of-Arrival~\cite{apsUsingAoA}, Time-of-Flight~\cite{sail}, SignalSLAM~\cite{signalSLAM}, or Inertial Sensing~\cite{inertialSensors}.

Given the challenges, we adopt an offline fingerprint-based approach to compute each device's location. Fingerprints have been demonstrated to be more accurate than model-based approaches in densely crowded spaces~\cite{FPvsmodel} and hence widely preferred.  
Our localization software processes the RSSI updates using well-known ``classical'' fingerprint-based technique~\cite{radar}.
Given the wide usage of this approach, our experiences and results apply to a majority of the localization algorithms.

Our primary contribution is to detail the cases where such a conventional approach succeeds and where it fails.  We highlight the related challenges for making the approach work in current, large-scale WiFi networks, and then develop appropriate solutions to overcome the observed challenges.
We collect three weeks of detailed ground truth data ($\approx 200$ landmarks) in our large-scale deployment, carefully construct a set of experimental studies to show two unique challenges -- \emph{Cardinality Mismatch} and \emph{High Client Scan Latency} associated with a server-side localization approach.
The three weeks of data is representative of our four years of data.

($a$) \emph{Cardinality Mismatch:} We define cardinality as the set of APs reporting for a client located at a specific landmark.
We first show that the cardinality, during the online phase, is \emph{often} quite different from the cardinality in the offline phase.
Note that this divergence is in the \emph{set} of reporting APs, and not just merely a mismatch in the values of the RSSI vectors. 
Intuitively, this upends the very premise of fingerprint-based systems that the cardinality seen at any landmark is the same during the offline and online phases. This phenomenon arises from the dynamic power and client management performed by a centralized controller in all commercial grade WiFi networks (for example, those provided by Aruba, Cisco, and other vendors) to achieve outcomes such as ($i$) minimize overall interference (shift neighboring APs to alternative channels), ($ii$) enhance throughput (shift clients to alternative APs), and ($iii$) reduce energy consumption (shut down redundant APs during periods of low load). 

($b$) \emph{High Client Scan Latency:} Most localization systems use client-side localization techniques where clients actively scan the network when they need a location fix. Specialized mobile apps trigger active scans at the client devices. However, when using server-side localization, the location system has no way to induce scans from client devices.This is because such a localization system is deployed in public places, e.g. universities and malls, where changes at the client device are not feasible. Hence, the system can only ``see'' clients when clients scan as part of their normal behavior.  However, as we show in
Section~\ref{sec:challenges}, most clients today, by default, do not prefer to scan. We analyze different usage mode of the devices. Our observations reveal that while mobile clients trigger scans only while handover, stationary clients trigger scans only when their screen is turned on.

These phenomena do not exist in small-scale deployments often used in the past pilot studies, where each AP is configured independently.
In large-scale deployments, where it is fairly common to use controller-managed WLANs with a large number of devices, these phenomena invariably persist to a great extent. To exemplify, we noticed $57.30$\% instances of cardinality mismatch in $2.4$ GHz and $30.60$\% in $5$ GHz in our deployment. We saw $90^{th}$\%ile of client scan interval to be $20$ minutes.  While localizing with fingerprint-based solutions in such environments, these phenomena translate to either \emph{minimal} or even worse \emph{no} matching APs, resulting in substantial delays between client location updates and ``teleporting'' of clients across the location. 
  
It is important to note that not only the schedule of these algorithms is non-deterministic but also their distribution during offline and online phases. This is attributed to the fundamental fact that the dynamics of WiFi networks such as load and interference, is non-deterministic in most of the cases and that the controller algorithm is a black-box to us.
Furthermore, the differences in signal propagation and scanning behavior of $2.4$ and $5$ GHz contribute to these problems.
We believe that we are the first to present the challenges of server-side localization as well as their mitigation. Our proposals are device-agnostic, simple, and easily integrable with any large-scale WiFi deployment to efficiently localize devices.

\noindent \textbf{Key Contributions:}

\begin{itemize}
\item We identify and describe a couple of novel and fundamental problems
associated with a server-side localization framework.  In particular, we
provide evidence for the (i) ``Cardinality Mismatch'' and (ii) ``High Client
Scan Latency'' problems, explain why these problems are progressively
becoming more significant in commercial WiFi deployments. We discuss the reasons why these problems are non-trivial to be solved given the challenge of no client/infrastructure-side allowed. Our entire analysis is for both frequency bands -- $2.4$ and $5$ GHz.
\item We provide valuable insights about the causes of these problems with extensive evaluations based on the ground truth data collected over three weeks for $200$ landmarks. Motivated from the real-world usage of clients, we study their 4 states -- Disconnected, Inactive, Intermittent, and Active. For each of these states, we analyze -- (i) their scanning behavior while being mobile as well as stationary and (ii) the impact of these states on the achieved cardinality in the presence and the absence of scans. We demonstrate the impact of considering ``only'' scanning frames as compared to ``only'' non-scanning frames on the localization errors. Furthermore, we compare our results with real-world scenario of having a mix of both categories of frames.
\item We propose heuristics to improve the accuracy of the localization in the face of these problems. We see an improvement from a minimum of $35.40$\% to a maximum of $100$\%. We show an improvement 
in the higher percentiles over SignalSLAM~\cite{signalSLAM}. This shows that our lessons learned have the potential of improving the existing localization algorithms. 
\item We describe our experiences with deploying, managing, and improving a
fingerprint-based WiFi localization system, which has been operational,
since $2013$, across the entire campus of Singapore Management University.  We not
only focus on the final ``best solution'' that uses RTLS data feeds, but also
discuss the challenges and pitfalls encountered over the years. 
\end{itemize}

\noindent \textbf{Paper Organization:} We discuss the related works in Section~\ref{sec:related_works}. We present the system architecture and the details of data collection in Section~\ref{sec:system-details}. We introduce the challenges, their evidences, and propose the solutions in Section~\ref{sec:challenges}. We discuss the challenges of localizing clients in real world deployments and the limitations of our proposed solutions in Section~\ref{sec:discussion}. We conclude in Section~\ref{sec:conclusion}.

\section{Related Work}
\label{sec:related_works}

In this section, we discuss existing solutions and their limitations for indoor localization.

\textbf{Fingerprint vs. Model-based Solutions:} One of the oldest localization techniques use either a fingerprint-based~\cite{minimizingcalibration, radar, surroundSense, monalisa, horus, zee, littleHumanIntervention, pushTheLimit} or model-based~\cite{localizeWithoutPreDeployment, zeroConfig, wigem, selfCalibratingLocalization} approach, or a combination of both~\cite{diversityInLocalization}. Overall, fingerprint-based solutions tend to have much higher accuracies than other approaches albeit with a high setup and maintenance cost~\cite{FPvsmodel}.
The fingerprint-based approach was pioneered by Radar~\cite{radar} and has spurred numerous follow-on research. For example, Horus~\cite{horus} uses a probabilistic technique to construct statistical radio maps, which can infer locations with centimeter level accuracy. PinLoc~\cite{monalisa} incorporates physical layer information into location fingerprints. Liu \textit{et al.}~\cite{pushTheLimit} improved accuracy by adopting a peer-assisted approach based on p$2$p acoustic ranging estimates. Another thread of research in this line is to reduce the fingerprinting effort, for example, using crowdsourcing~\cite{littleHumanIntervention, will, zee, localizationUsingPriorInfo} and down-sampling~\cite{minimizingcalibration}.
The mathematical signal propagation model approach~\cite{wigem, localizeWithoutPreDeployment} has the benefit of easy deployment (no need for fingerprints) although its accuracy suffers when the environment layout or crowd dynamics change~\cite{selfCalibratingLocalization}. Systems, such as EZ, improve the accuracy by additionally using GPS to guide the model construction.

\textbf{Client vs. Infrastructure-based solutions:} There is a rich history of client-based indoor location solutions, to name a few, SignalSLAM~\cite{signalSLAM}, SurroundSense~\cite{surroundSense}, UnLoc~\cite{unsupervisedIndoorLoc}, and many others ~\cite{centaur, will, zee, localizeWithoutInfra, virtualCompass, placeLab}. All of them share some commonalities in that they extract sensor signals (of various types) from client devices to localize. The location algorithms usually run on the device itself; however, it is also possible to run the algorithm on a server and use the signals from multiple clients to achieve better performance~\cite{pushTheLimit}. Overall, client-based solutions have very high accuracy (centimeter resolution in some cases~\cite{horus, monalisa}). 
An alternative would be to pull signal measurements directly from the WiFi infrastructure, similar to what our solution does. The research community has only lightly explored this approach since it requires full access to the WLAN controllers, which is usually proprietary. Our main competitors are the commercial WiFi providers themselves. In particular,
both Cisco~\cite{cisco} and Aruba~\cite{aruba} offer location services. These solutions use server-side tracking coupled with model-based approaches (to eliminate fingerprint setup overhead).

\textbf{Other Solutions:} There are several other solutions, complementary to the signal strength-based technique. Time-based solutions~\cite{pinpoint, sensorToA, caesar, sail, cupid} use the arrival time of signals to estimate the distance between client and AP, while angle-based solutions~\cite{apsUsingAoA, zeroStartupCosts, phaser, arraytrack, spotFi} utilize angle of arrival information, estimated from a antenna array, to locate mobile users.  Recently, the notion of passive location tracking~\cite{wifiImaging, humanActivity, reusing60GHz, seeThroughWalls, mtrack} has been proposed, which does not assume people carry devices. In large and crowded venues, however, the feasibility and accuracy of such passive tracking is still an open question. Other systems like light-based localization~\cite{epsilon, pharos, localizeUsingLights} and acoustic-based localization~\cite{swordfight, audioCapture, sonoloc, acousticOccupany}.

\textbf{Limitations of above solutions:} These solutions can achieve higher accuracy, but they have at least one of the following limitations -- ($a$) need of a customized hardware, which cannot be implemented in large-scale deployments, ($b$) installation of client application, making them hard to scale, ($c$) rooting client OS - Android or iOS, which limits their generalizability, ($d$) energy savvy, ($e$) high error rates in dense networks, and ($f$) proprietary and expensive to deploy (especially, solutions from vendors like Cisco and Aruba).

To summarize, even though several wonderful solutions are available, their scalability is still a question.
Therefore, we advocate using server-side localization approach with fingerprints.
Our aim is not to compare the efficacy of different approaches, but to address the challenges of practical and widely deployed device-agnostic indoor localization using today's WiFi standards and hardware, for example, use of $5$ GHz band and controller-based architecture.
\section{System Architecture and Data Collection}
\label{sec:system-details}
In this section, we present the details about system architecture and the dataset.
\subsection{Background \& Deployment}
\label{subsec:background}

This work began in $2013$ when we started deploying a WiFi-based localization
solution across the entire campus. 
It has since gone through much major and minor
evolutions.
However, in this paper, we focus our
evaluation and results on just one venue -- a university, as we have
full access to that venue.

Our university campus has seven schools in different buildings.  
Five buildings have six floors, remaining two have five and three floors
respectively, with a floor area of $\approx 70,000$ $m^2$. 
Landmarks, characterized by water sprinklers are deployed every three meters, on a
given floor denote a particular location.  There are $3203$ landmarks across
thirty-eight floors of seven schools.  WLAN deployment includes $750$+ dual-band APs,
centrally controlled by eleven WiFi controllers, with $\approx 4000$
associated clients per day.
\begin{figure}[t]
\centering
\includegraphics[scale=0.45]{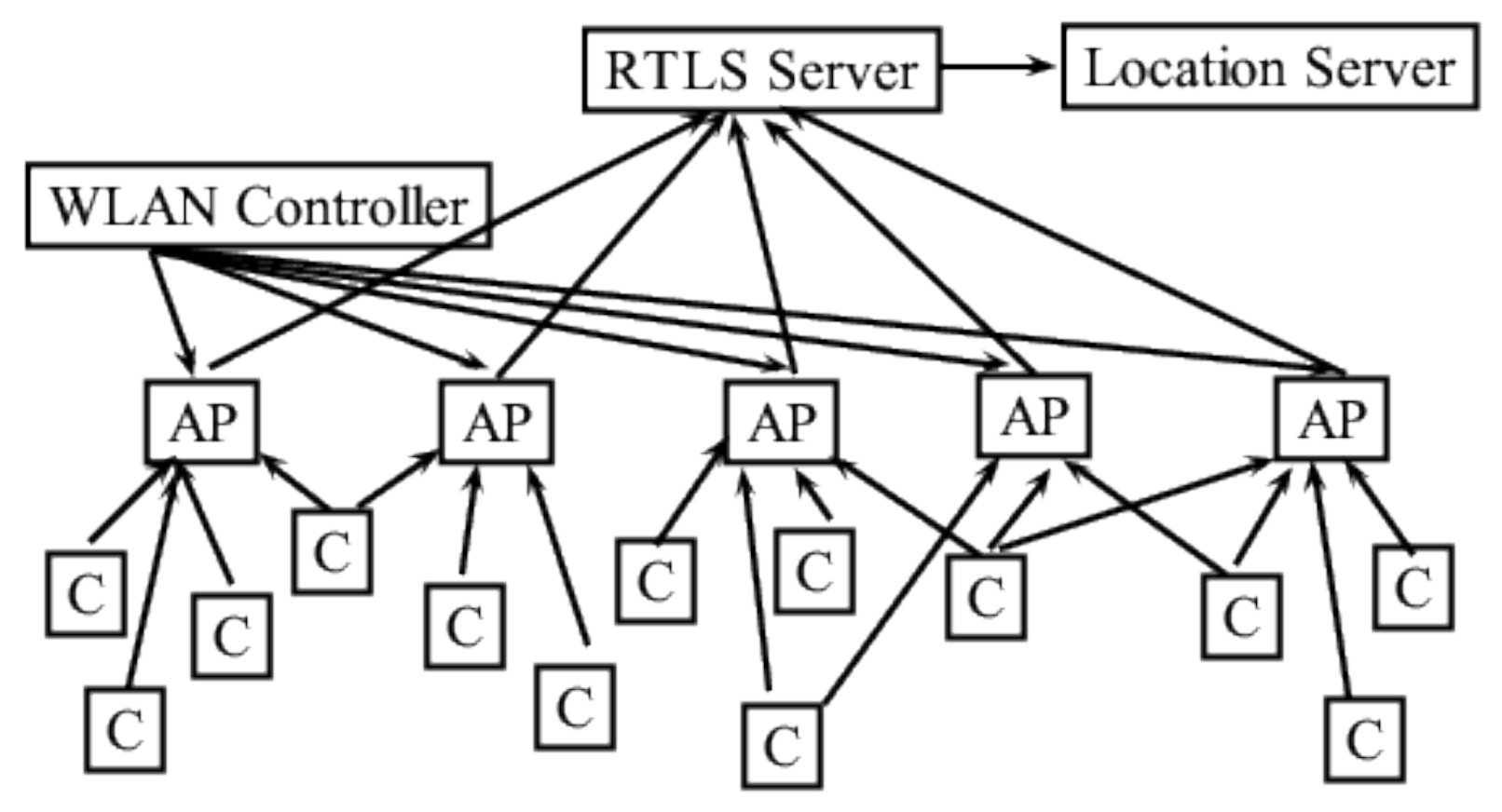}
\caption{Block diagram of Indoor Localization System. Note that lines between AP and C denote the coverage area of AP and not the association. \textit{Legend: AP - Access Point, C - Client, WLAN - Wireless LAN, RTLS - Real-Time Location System}} 
\label{Fig:LocationSystem}
\end{figure}

\begin{table}[t]
\centering
\begin{tabular}{@{} lm{8cm} @{}}
\toprule
\textbf{Field} & \textbf{Description} \\
\hline
Timestamp & AP Epoch time (milliseconds) \\
Client MAC & SHA$1$ of original MAC address \\
Age & \#Seconds since the client was last seen at an AP\\
Channel & Band ($2.4$/$5$ GHz) on which client was seen \\
AP & MAC address of the Access Point \\
Association Status & Client's association status (associated/unassociated) \\
Data Rate & MAC layer bit-rate of last transmission by the client\\
RSSI & Average RSSI for duration when client was seen \\
\bottomrule
\end{tabular}
\caption{Details of RTLS data feeds}
\label{tbl:rtls_feeds}
\end{table}
\subsection{System Architecture}

Figure \ref{Fig:LocationSystem} represents the primary building blocks of the system.
The system is bootstrapped with APs configured by the WLAN controller to send RTLS data feeds every $5$ seconds to the RTLS server. 
Most commercial WLAN infrastructures allow such a configuration.
Once configured, APs bypass WLAN controller and report RTLS data feeds directly to our Location Server.
Table~\ref{tbl:rtls_feeds} presents all the fields contained in an RTLS data feed per client.
The reported RSSI value is not on a per-frame basis, but a summarized value from multiple received frames.
The Location Server analyzes these RTLS data feeds for the signal strengths reported by different APs to estimate the location of a client.
Note that the APs do not report the type of frames. 
They gather information from their current channel of operation and scan other channels to collect data.
Vendors have microscopic details of what APs measure~\cite{aruba_location}, however as an end-user we do not have access to any more information than what is specified. 
Nevertheless, even this information at large-scale gives us a view of the entire network from a single vantage point.

\subsection{Recording of the Fingerprints}
\label{subsec:fingerprinting}
We define a fingerprint as a vector of RSSI from APs for a given client.
We consider two types of fingerprints -- offline and online.
An offline fingerprint is collected and stored in a database before the process of localization is bootstrapped, while an online fingerprint is collected in real-time.

\textbf{Offline Fingerprinting} A two-dimensional offline fingerprint map is prepared for each landmark on the per-floor basis.
The client devices used for fingerprinting were dual-band Android phones, which were associated with the network, and they actively scanned for APs.
For each landmark, the device collected data for $5$ minutes. 
While the clients scan their vicinity, APs collate RSSI reports for the client and send their measurements as RTLS data feeds to the Location Server.
For a given landmark $L_{i}$, an offline fingerprint takes the following form:
\begin{equation}
<L_{i},B,AP_{1}:RSSI_{1};...;AP_{n}:RSSI_{n};>
\label{eq:fpmap_offline}
\vspace*{-0.05in}
\end{equation}
We maintain fingerprints for both $2.4$ and $5$ GHz frequency bands. 
Band $B$, in the above equation, takes a value of band being recorded. 
The vectors are stored in a database on the Location Server.

\textbf{Online Fingerprinting} Localization of a client is done with online fingerprints. An online fingerprint takes the same syntax as offline fingerprints in Equation~\ref{eq:fpmap_offline}, except the landmark, as shown below:
\begin{equation}
<B,AP_{1}:RSSI_{1};...;AP_{m}:RSSI_{m};>
\label{eq:fpmap_online}
\vspace*{-0.05in}
\end{equation}

Now, we match this online fingerprint with offline fingerprints of each landmark to calculate the distance in signal space, as discussed in~\cite{radar}. The landmark with minimum distance in signal space is reported as the probable location of the client.

\subsection{Pre-processing of the Data}
Now, we present the details of data collection and its processing.

\begin{figure}[t]
\centering
\includegraphics[scale=0.3]{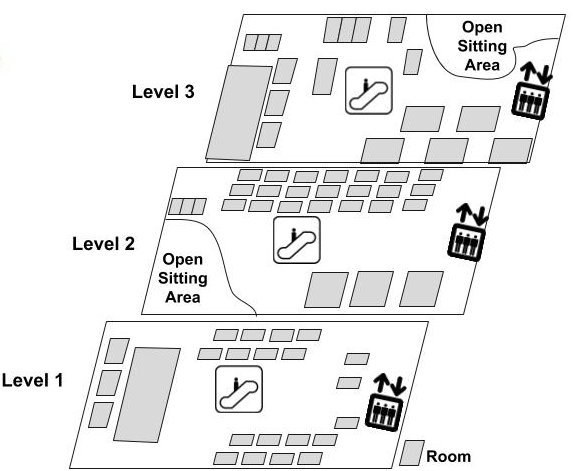}
\caption{Floor Map of the school where we collected ground truth data.}
\label{Fig:floormap}
\end{figure}

\subsubsection{\textbf{Collection of the Ground Truth}}
\label{subsubsec:gt_collection}

We collect the ground truth data for online fingerprints.
We want to correlate the data collection with real-world usage scenarios. 
Therefore, we choose four most common states of WiFi devices as per their WiFi association status and Data transmission. 
The states are -- ($i$) Disconnected, ($ii$) WiFi Associated -- ($ii.a$) Never actively used by user, ($ii.b$) Intermittently used, and ($ii.c$) Actively used. 
These states implicitly modulate the scanning frequency.
We use a separate phone for each state; thus, we use $4$ Samsung Galaxy S$7$ phones to record ground truth for each landmark.

State ($a$) client is disconnected. In this state, WiFi is turned on but not associated with any AP and screen remains off throughout the data collection. Therefore, only traffic generated from this client is scanning traffic and no data traffic. We ensure that this client did not follow MAC address randomization, which most latest devices follow in the unassociated state~\cite{MACRandomization}.
State ($b$) client is associated but inactive. In this state, WiFi is turned on, it is associated but the screen remains off throughout the data collection. 
State ($c$) client is associated and intermittently used. In this state, WiFi is on, the client is associated and user intermittently uses the device. This is one of the most common state for mobile devices and previous research~\cite{sniffer-based-inference} states scanning is triggered whenever screen of the device is lit up. 
State ($d$) client is associated and actively used. In this state, WiFi is on, the client is associated, and a YouTube video plays throughout the data collection. This state generates most data traffic, \textit{i.e.} non-scanning traffic. 
Each client stayed at a landmark for about a minute before it moved to the next landmark. We manually noted down start time and end time for every landmark at the granularity of seconds.
We did this exercise for $3203$ landmarks of our university, collected $86$ hours worth of data, that accounts for $54,096$ files carrying $274$ GB of data. 
The amount of time to localize a client is $40$ seconds.
Processing the entire dataset would take $\approx 100$ days. 
Therefore, given the size of the entire dataset, we present our analysis of $200$ landmarks, which accounts for $3121$ files with $15.3$ GB of data. 
Figure~\ref{Fig:floormap} shows the floor map of one of the schools whose data we refer for our analysis.

Our aim is to demonstrate the challenges associated with fingerprint\hyp{}based localization. These challenges apply to all the solutions that employ fingerprint-based localization, irrespective of the type of device present in the network. The variation of RSSI with device heterogeneity is well known~\cite{RSSIVariation} and that will further exacerbate the problems identified by this paper. 
We collect ground truth with only one device so that we can highlight issues without any complications added by heterogeneous devices.

\subsubsection{\textbf{Pre-processing of the RTLS Data Feeds}}
Our code reads every feed to extract the details of APs reporting a particular client.
RTLS data feeds, may obtain stale records for a client. Therefore, we filter the raw RTLS data feed for the latest values, with age less than or equal to $15$ seconds, and the RSSI should be greater than or equal to $-72$ dBm. The threshold for age is a heuristic to take the most recent readings. The threshold for RSSI is decided based on the fact that a client loses association when RSSI is below $-72$ dBm. 

For our analysis, we classify MAC layer frames in two classes ($a$) \emph{Scanning Frames} -- high power and low bit rate probe requests and ($b$) \emph{Non-Scanning Frames} -- all other MAC layer frames.
Offline fingerprints are derived from the scanning frames, which are known to provide accurate distance estimates as they are transmitted at full power. 
In the offline phase, a client is configured to scan continuously. However, in the online phase we have no control over the scanning behavior of the client, resulting in a mix of scanning and non-scanning frames. Therefore, while localizing with the fingerprints, RSSIs available for matching are from the different categories of frames.
RTLS data feeds do not report the type of frame and do not have a one-to-one mapping of MAC layer frames to the feeds.
Therefore, we devise a probabilistic approach to identify these frames.

We design with a set of controlled experiments, where we configured the client in one of the two settings at a time ($a$) send scanning frames only and ($b$) send non-scanning frames only. These two settings are mutually exclusive. 
We collected the traffic from the client with a sniffer as well as the corresponding RTLS data feeds.
Then, we compare both the logs -- sniffer and RTLS, to confirm the frame types and the corresponding data rates.

Our analysis reveals that when a client is associated and sending non-scanning frames, 
the AP to which it is associated reports the client as \emph{associated}. 
The data rates of the RTLS data feeds vary among various $802.11g$ rates, e.g. $1$, $2$, $5.5$, ...,$54$ Mbps.
Even though, our network deployment is dual-band and supports the latest $802.11$ standards including $802.11ac$, still the rates reported in the RTLS data feeds follow $802.11$g.
We do not have any visibility in the controller's algorithm to deduce the reason for this mismatch in the reported data rates.
However, when a client sends scanning frames, all the APs that could see the client report the client as \emph{unassociated} 
and the data rates reported is fixed at either $1$, $6$, or $24$ Mbps, as per the configured probe response rate.

We use these facts to differentiate non-scanning and scanning RTLS data feeds. 
We believe this approach correctly infers scanning frames because ($a$) the data rates are fixed to $1$, $6$, or $24$ Mbps, ($b$) when an associated client scans, other APs report that client as unassociated, and ($c$) an unassociated client can only send either scanning or association frames. 
However, our approach may still incorrectly identify a scanning frame as non-scanning in the following cases -- ($a$) When an associated client scans and the AP, to which it is associated, reports. This AP reports the client as associated and its data rate as $1$, $6$, or $24$ Mbps. In this case, these rates may also be because of the non-scanning frames. 
We identify such feeds as non-scanning.
($b$) When an unassociated client sends association or authentication frames. In this case also, the rates overlap with the scanning data rates and the association status is reported as unassociated. Here, we incorrectly identify non-scanning frames as scanning frames.
However, these cases are rare.
For other cases, our approach is deterministically correct.

\section{Challenges Discovered}
\label{sec:challenges}

In this section, we give evidence of the issues, namely Cardinality Mismatch and High Client Scan Latencies.
We compare the severity of these issues for both the frequency bands, identify the causes behind these issues, and
measure their impact on the issues.

\begin{figure*}[t]
\centering
        \begin{subfigure}[b]{0.46\textwidth}
        \centering
		\includegraphics[scale=0.5]{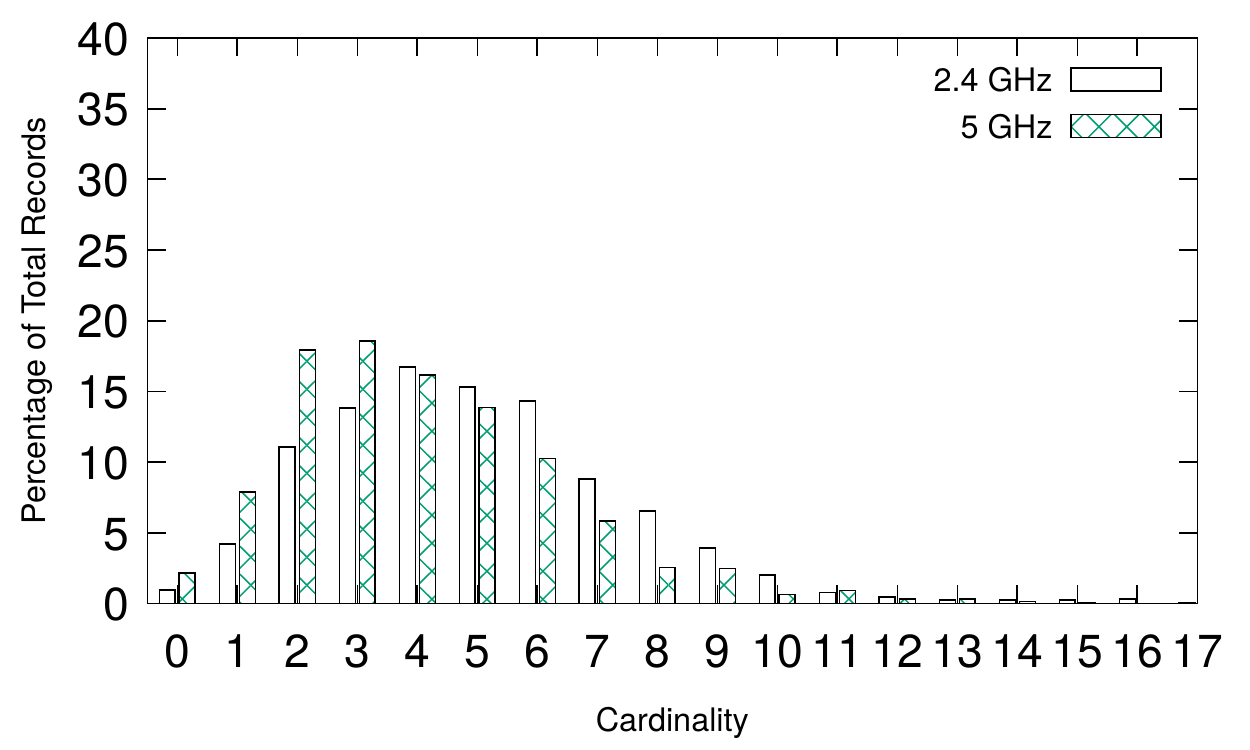}
                \caption{Offline Phase }
                \label{Fig:FPMAP-Cardinality}
        \end{subfigure}%
        \qquad
        \begin{subfigure}[b]{0.46\textwidth}
        \centering
		\includegraphics[scale=0.5]{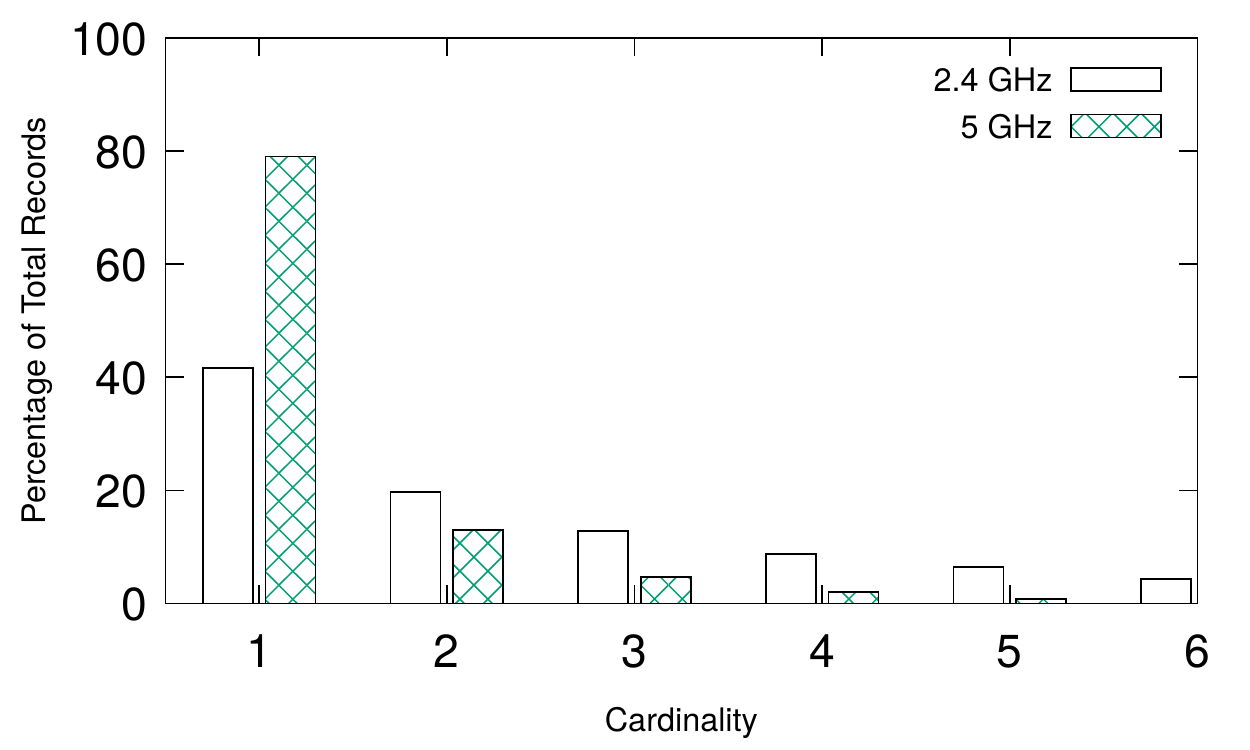}
                \caption{Online Phase}
                \label{Fig:ReportingAPs}
        \end{subfigure}
\caption{Cardinalities observed during the offline and online phases. For both the phases, the cardinalities are lower for $5$ GHz. During the online phase, there is a substantial decrease in the cardinality for both bands as compared to the offline phase.}\label{Fig:cardinality_issues}
\end{figure*}

\begin{figure*}[t]
\centering
        \begin{subfigure}[b]{0.46\textwidth}
        \centering
		\includegraphics[scale=0.5]{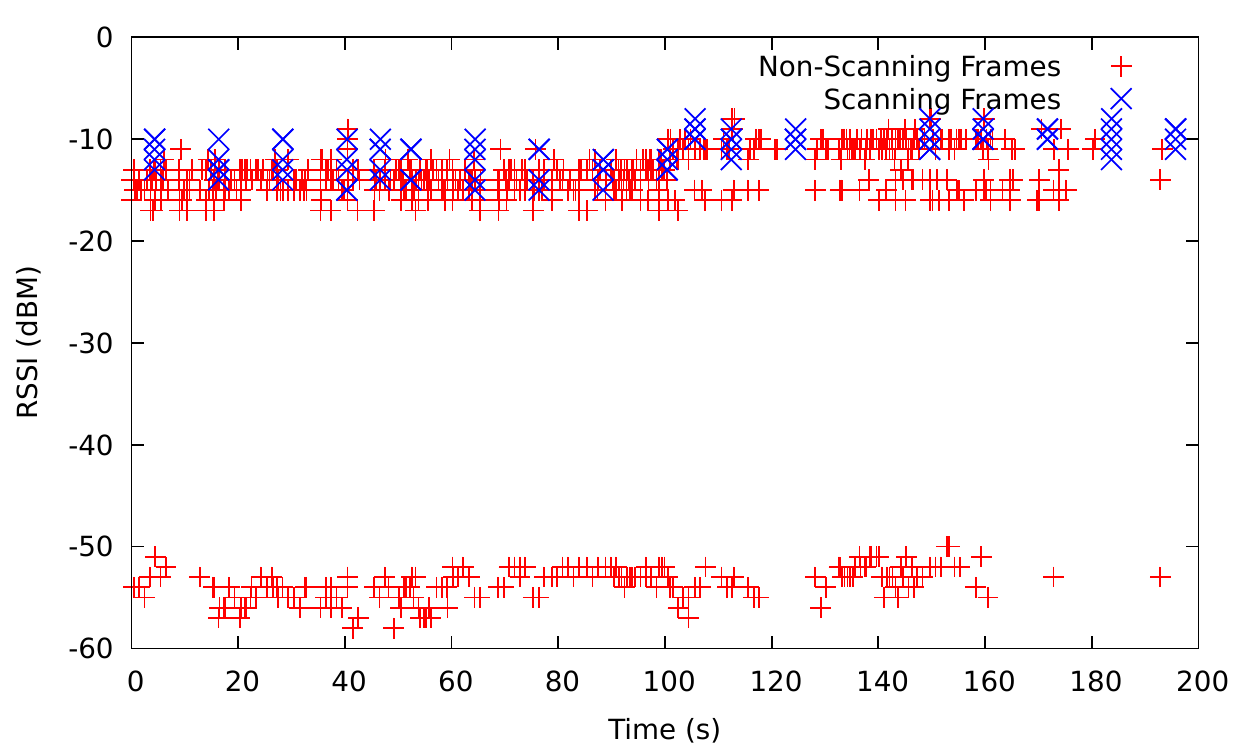}
                \caption{Client Close To AP}
                \label{Fig:ClientCloseToAP-ScanVsNoScan}
        \end{subfigure}%
        \qquad
        \begin{subfigure}[b]{0.46\textwidth}
        \centering
		\includegraphics[scale=0.5]{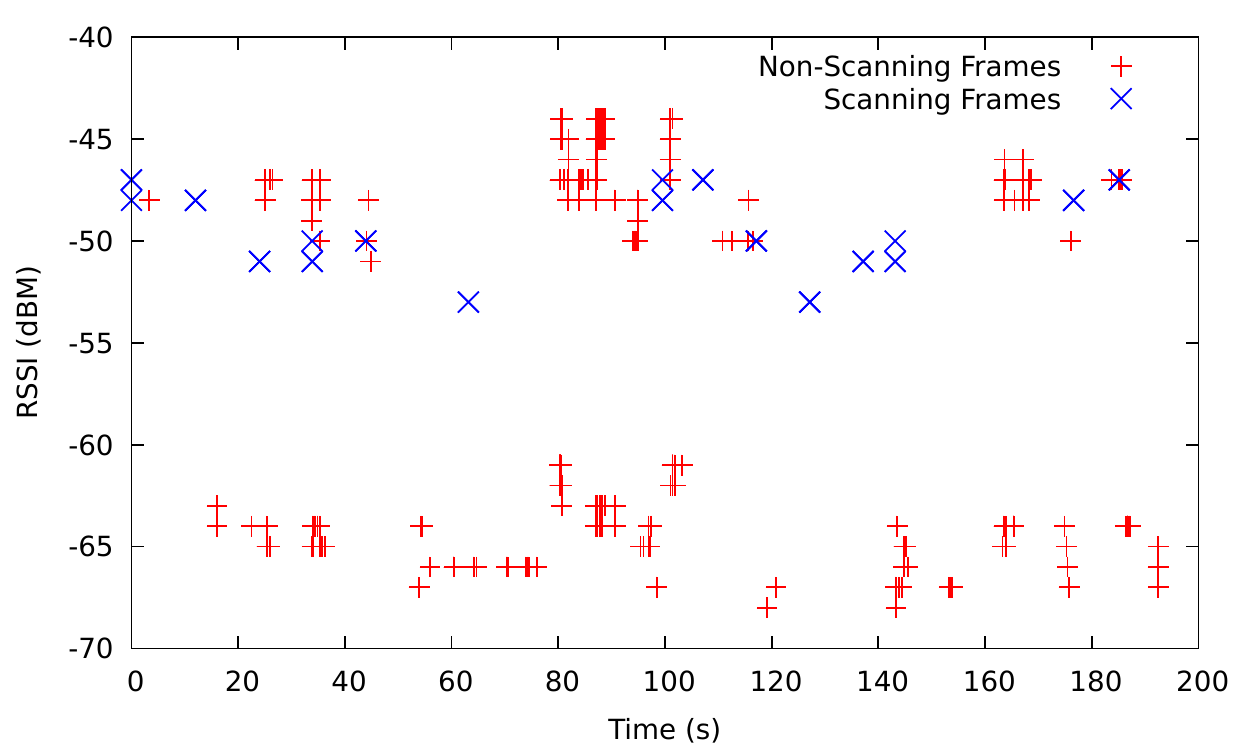}
                \caption{Client Far From AP}
                \label{Fig:ClientFarFromAP-ScanVsNoScan}
        \end{subfigure}%
\caption{Variations in RSSI for scanning and non-scanning frames in two scenarios -- ($a$) client close to the AP and ($b$) client far from the AP.  For both cases the RSSI from scanning frames vary far lesser than the non-scanning frames.}\label{Fig:rssi_mismatch_scanvsnoscan}

\end{figure*}

\begin{figure}[t]

\centering
\includegraphics[scale=0.6]{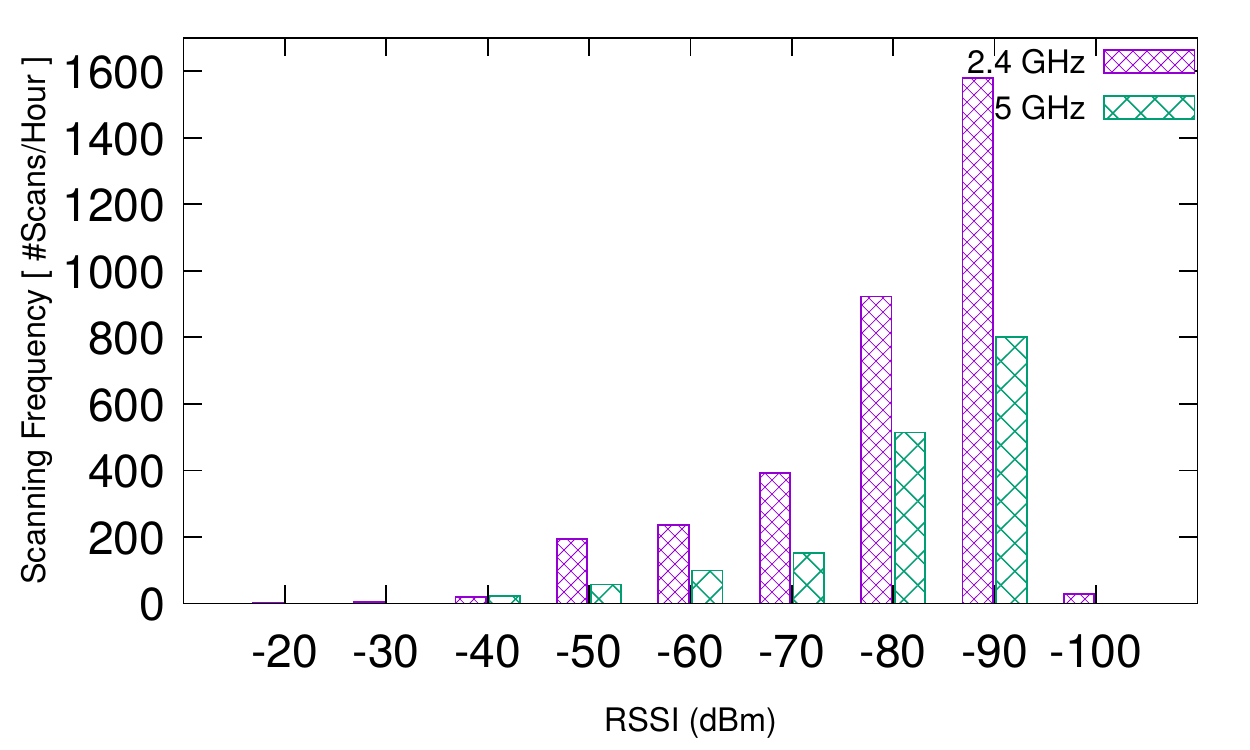}

\caption{Frequency of scanning in both band increases as RSSI reduces. $2.4$ GHz experiences higher scan frequencies.}

\label{Fig:Hist-RSSIs-ScanFreq}

\end{figure}
\subsection{Evidence of the Issues}
The Cardinality Mismatch arises from the dynamic power and client management performed by a centralized controller as well as the client-side power management.
Given the dynamic nature of these management policies, 
it is not possible to estimate their implications on the Cardinality Mismatch,
and thereby on the localization errors.
We take an empirical approach to see whether 
($a$) we can find out the severity of these implications on the Cardinality Mismatch and the localization error
and ($b$) identify the tunability of the implicating factors.

Figure~\ref{Fig:cardinality_issues} plots the differences in cardinality 
between the offline and online phases for $2.4$ and $5$ GHz.
Figure~\ref{Fig:FPMAP-Cardinality} shows the cardinality observed in our offline fingerprints.
Figure~\ref{Fig:ReportingAPs} shows the cardinality observed during the online phase. 
While the maximum cardinality is $16$ during the offline phase, it is merely $6$ in the online phase.
This shows the spectrum of the Cardinality Mismatch.
In the online phase, $80$\% of the time only $1$ AP reports for a client in $5$ GHz while $40$\% in $2.4$ GHz.
Any fingerprint-based algorithm will be adversely affected by such a big difference in the cardinality.
For each band, we find out how much is the extent of the Cardinality Mismatch.
Overall, across all the cardinalities, $2.4$ GHz has $57.30\%$ mismatches and $5$ GHz has $30.6\%$ mismatches. 
The $5$ GHz band is more adversely affected by the Cardinality Mismatch issue because it experiences lower cardinality, which increases the chances of a mismatch.
$2.4$ GHz always sees higher cardinality than $5$ GHz, both during the offline and online phases. 
This is because -- ($a$) signals in $2.4$ GHz travel farther than that of $5$ GHz, and ($b$) 
the number of scanning frames transmitted in $2.4$ GHz.
Unlike the data frames, the scanning frames are broadcasted 
and hence heard by more number of APs.
As the number of scanning frames increases, more APs hear them and revert,
thereby increasing the cardinality.

Besides, the RSSI variation for the scanning frames is lesser compared to that of the data frames.
To validate, we perform a controlled experiment with a stationary client and collect client's traffic using a sniffer. The client has an ongoing data transmission and periodic scanning is triggered every $15$ seconds. From the sniffed packet capture, we extract per-frame RSSI. The experiment is repeated for two scenarios- ($a$) the client is close to the AP and ($b$) the client is far from the AP. 
With these two scenarios, we emulate the client behavior for low and high RSSI from the AP.

Figure~\ref{Fig:rssi_mismatch_scanvsnoscan} shows the RSSI measurements in two cases.
In the first scenario, when the client is close to an AP, RSSI of the scanning frames varies by up to $10$ dB and for non-scanning frames it varies by up to $50$ dB. Similarly, in the second scenario, when the client is far from AP, RSSI of scanning frames varies by up to $5$ dB and for non-scanning frames it varies by up to $30$ dB.
Both our experiments validate that the RSSI from scanning frames vary far lesser than the non-scanning frames. 
This means the online RSSI from scanning frames match more closely 
and is a much more reliable indicator of the client's position.
We want to study how clients in their default configuration behave in \emph{real} networks; therefore we do not modify the default behavior of client driver in any way.
We repeated the experiment with devices of Samsung, Nexus, Xiaomi, and iPhone.

Next, we study the effect of the band on the frequency of scanning.
We collect WiFi traffic with sniffers listening on the channels in operation at that time in both the bands for $6$ hours. Data from $200$ WiFi clients is recorded.
Figure~\ref{Fig:Hist-RSSIs-ScanFreq} shows the plot.
For both $2.4$ and $5$ GHz bands, the frequency increases as RSSI reduces. 
Overall, the frequency is lesser for $5$ GHz, even though most ($\approx 2X$) of the clients in our network associate in $5$ GHz.
More the frequency of scanning, lesser is the chance of Cardinality Mismatch.
Our comparative analysis of the two bands revealed that the instance of frame losses and poor connection quality, which cause scanning, are much lower in $5$ GHz due to lower interference. 
The analysis of the scanning behavior of our clients reveals that -- ($a$) $90^{th}$ \%ile values of scanning intervals is in the order of few $1000$ seconds, which is a lot for fingerprint based solutions, ($b$) $5$ GHz is the least preferred band of scanning, and ($c$) clients rarely scan both the frequency bands.
Hence, we rule out the possibility of the reduced range of $5$ GHz resulting in lesser scanning frames.

\subsection{Causes Behind the Issues}

Next, we study the combined effect of frequency of scanning, \textit{i.e.}, number of scans per hour, and transmission distance on the cardinality.
For this, we consider clients configured in one of the four states as discussed in Section~\ref{subsubsec:gt_collection}.
Note that each state implicitly controls the amount of scanning. We do not manually control scanning behavior to imitate the real-world.

\begin{figure}[t]
\centering
\includegraphics[scale=0.6]{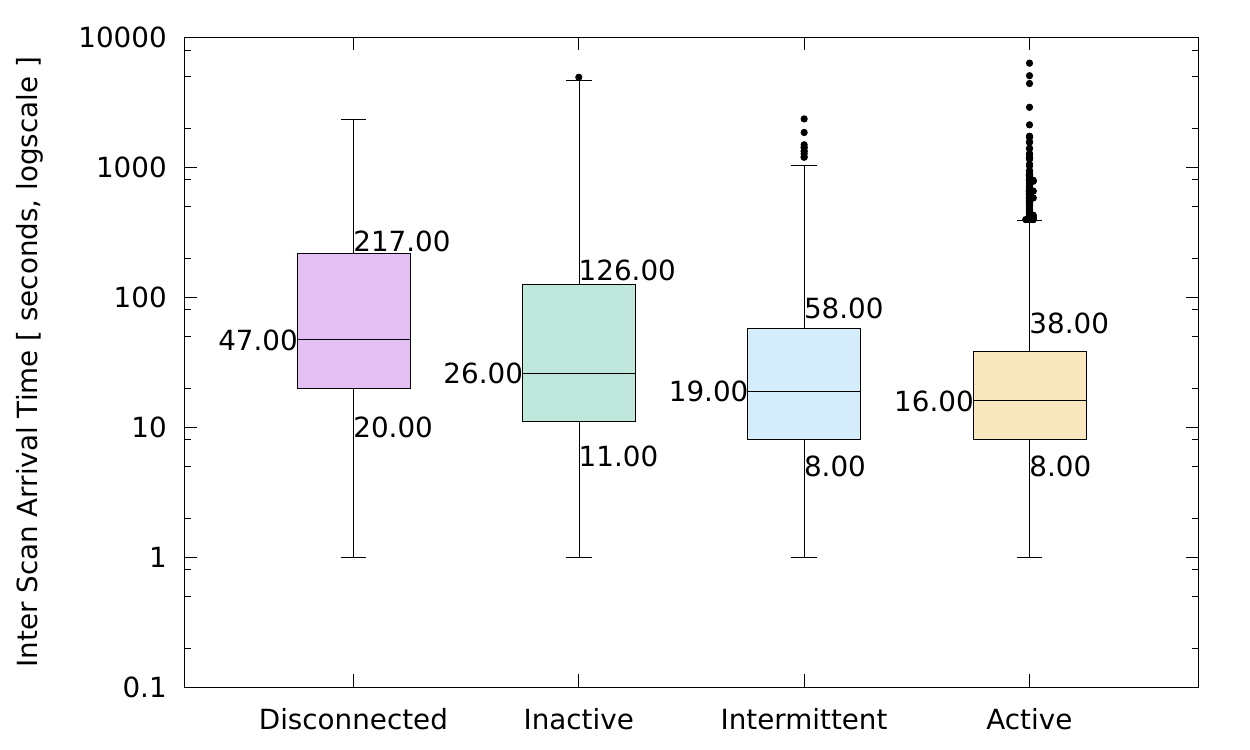}
\caption{Inter Scan Arrival Time for a client in four states -- Disconnected, Inactive, Intermittent, and Active. 
The intermittent and active states have median scanning time of $16$-$19$ seconds, while it increases to $26$-$47$ seconds for the inactive and disconnected clients. Notice that the upper quartile inter scan arrival times go up to $1000$ of seconds, which means clients mostly do not scan and hence the reduced cardinality.}
\label{fig:ISAT}
\end{figure}

Figure~\ref{fig:ISAT} shows the interval between two consecutive scans, \textit{i.e.}, the inter-scan arrival time, observed for each client state when the client moves from one landmark to another. We find that the median scan intervals are $15$-$20$ seconds for a client with either WiFi intermittently used or continuously used, while it is $26$-$47$ seconds for a client whose WiFi is disconnected or not actively used. However, in all the cases, $90^{th}$ percentile values are in thousands of seconds, which signify that clients scan infrequently in reality.

\begin{table*}[h!]
\centering
\small
\begin{tabular}{m{4cm}m{2.4cm}m{1.7cm}m{3cm}m{1cm}}
\toprule
\textbf{Phone} & \textbf{Disconnected} & \textbf{Inactive} & \textbf{Intermittent} & \textbf{Active}\\
\hline
iPhone $6$ (iOS $10.3$) & Random & \xmark & \xmark & \xmark\\
Nexus $5$X (Android $7.1$) & $100$-$300$s & \xmark & Screen Off$\rightarrow$On & Once\\
Galaxy S$7$ (Android $6.0$) & $100$-$2600$s & $1200$s & Screen Off$\rightarrow$On & \xmark\\
Galaxy S$3$ (Android $4.0$) & $240$s & $300$-$1300$s & Screen Off$\rightarrow$On & \xmark\\
Moto G$4$ (Android $7.1$) & $100$-$150$s & $10$-$1000$s & Screen Off$\rightarrow$On & $15$s\\
SonyXperia (Android $6.0$) & Once & $5$-$30$s & Screen Off$\rightarrow$On & Once\\
\bottomrule
\end{tabular}
\caption{Scanning Behavior of the Stationary clients. Most devices either do not trigger scans or perform minimal scans while in Disconnected, Inactive, or Active states. It may take up to 2600 seconds for devices to a trigger scan. This latency negatively reduces the localization accuracy. An exception is the Intermittent state in which all the Android devices trigger scans as and when the screen is turned on, otherwise the devices stay silent. The reduced number of active scans, across the states, result in inaccurate localization.}
\label{tbl:scan_behavior}
\end{table*}

We evaluate scanning behavior of stationary clients by closely monitoring $6$ different models of phone. Table~\ref{tbl:scan_behavior} summarizes the measurements and confirms our observations of the reduced scans. Unlike mobile clients, stationary clients tend to scan much lesser. The clients trigger scan almost every time when the screen is turned on while being in the intermittent state. This is especially true for the Android clients. However, scanning is infrequent in all the other states; even if the client is in the Active state. Such a behavior results in the reduced cardinality and therefore inaccurate localization.

\begin{figure*}
\scriptsize
\centering
\begin{tabular}{cc}
  \includegraphics[scale=0.5]{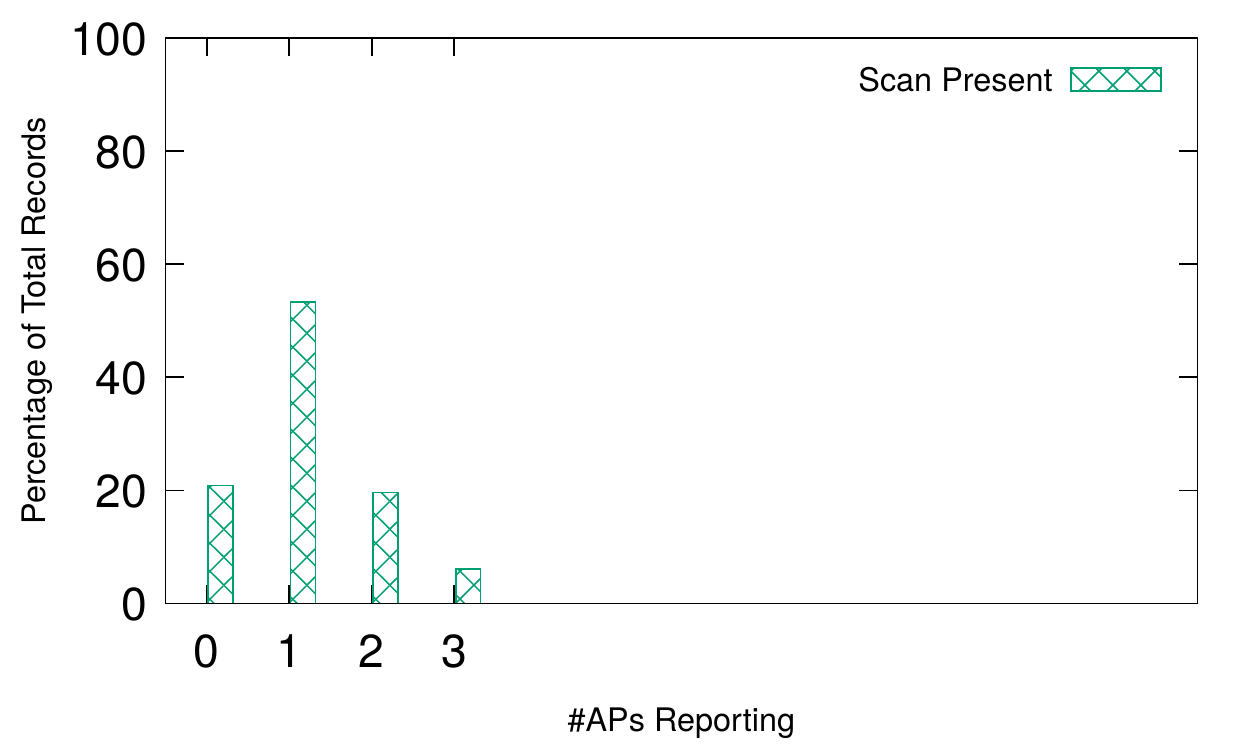}& \hspace{0.1in} \includegraphics[scale=0.5]{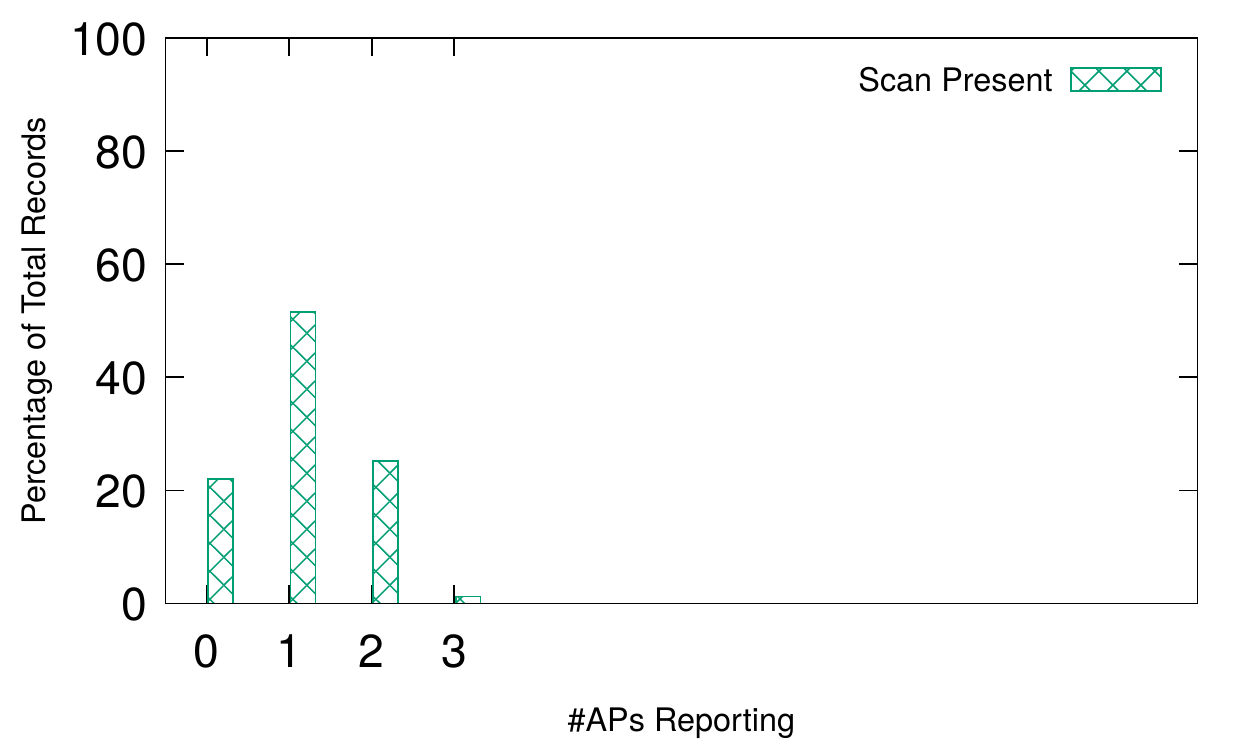}
\\(a) \textbf{State: Disconnected - $2.4$ GHz} & (b) \textbf{State: Disconnected - $5$ GHz}\\
  \includegraphics[scale=0.5]{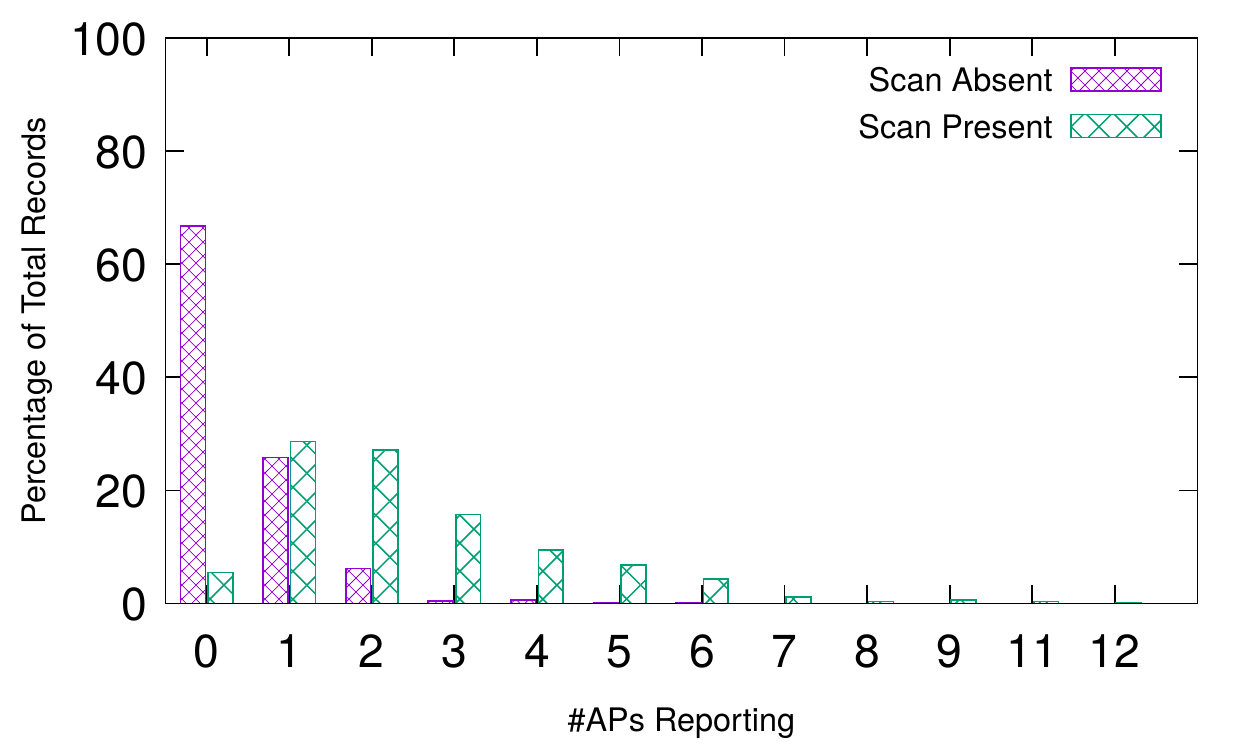}& \hspace{0.1in} \includegraphics[scale=0.5]{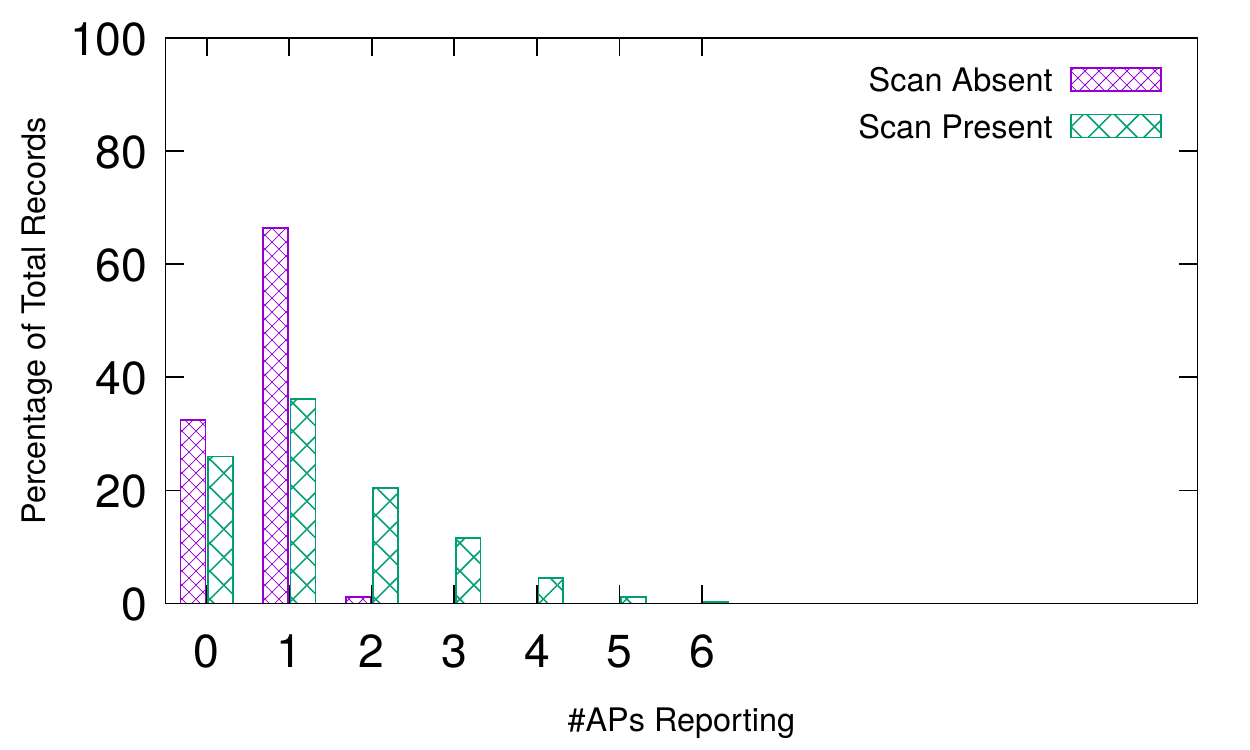}
\\(c) \textbf{State: Inactive - $2.4$ GHz} & (d) \textbf{State: Inactive - $5$ GHz} \\
  \includegraphics[scale=0.5]{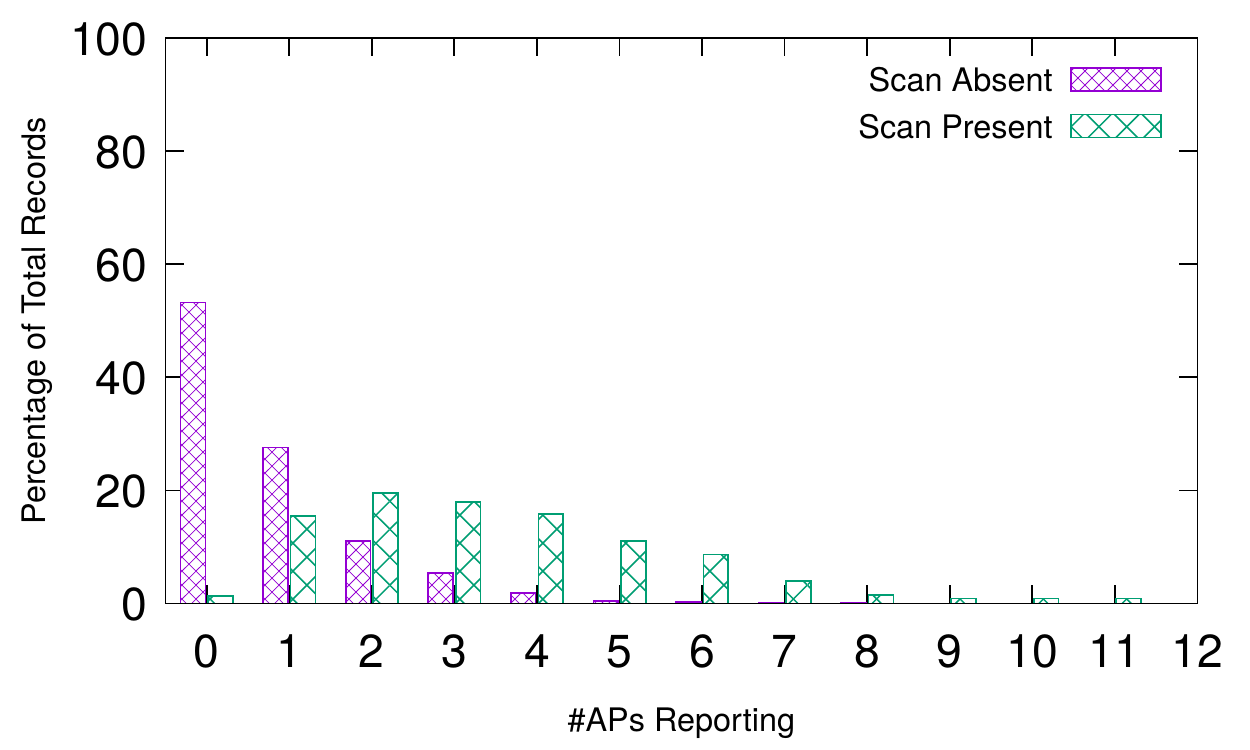}& \hspace{0.1in} \includegraphics[scale=0.5]{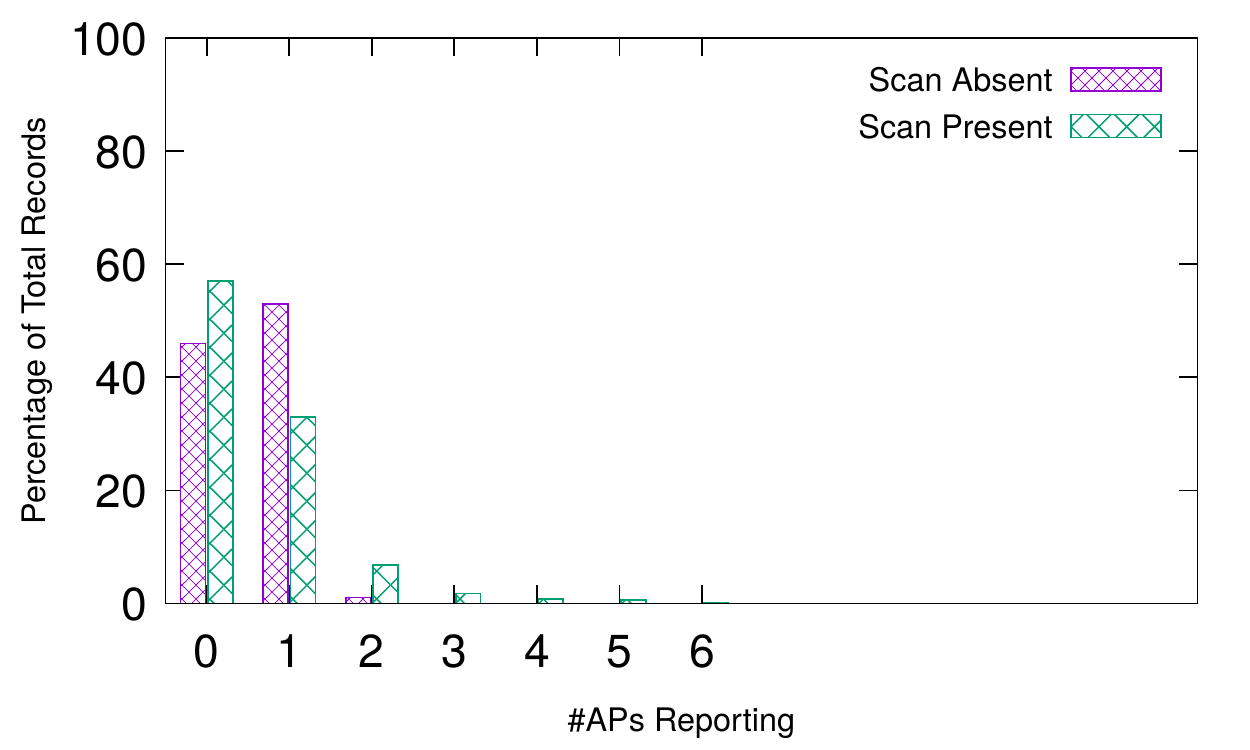}
\\(e) \textbf{State: Intermittent - $2.4$ GHz } & (f) \textbf{State: Intermittent - $5$ GHz} \\
  \includegraphics[scale=0.5]{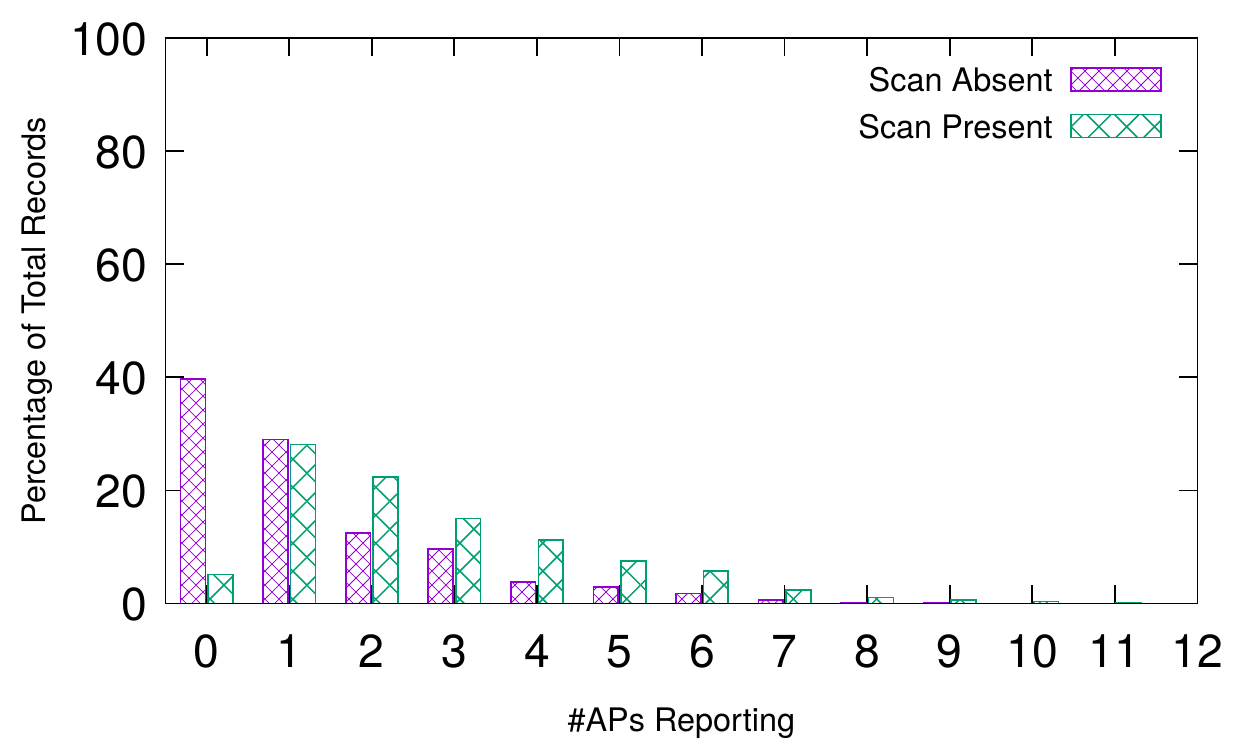}& \hspace{0.1in} \includegraphics[scale=0.5]{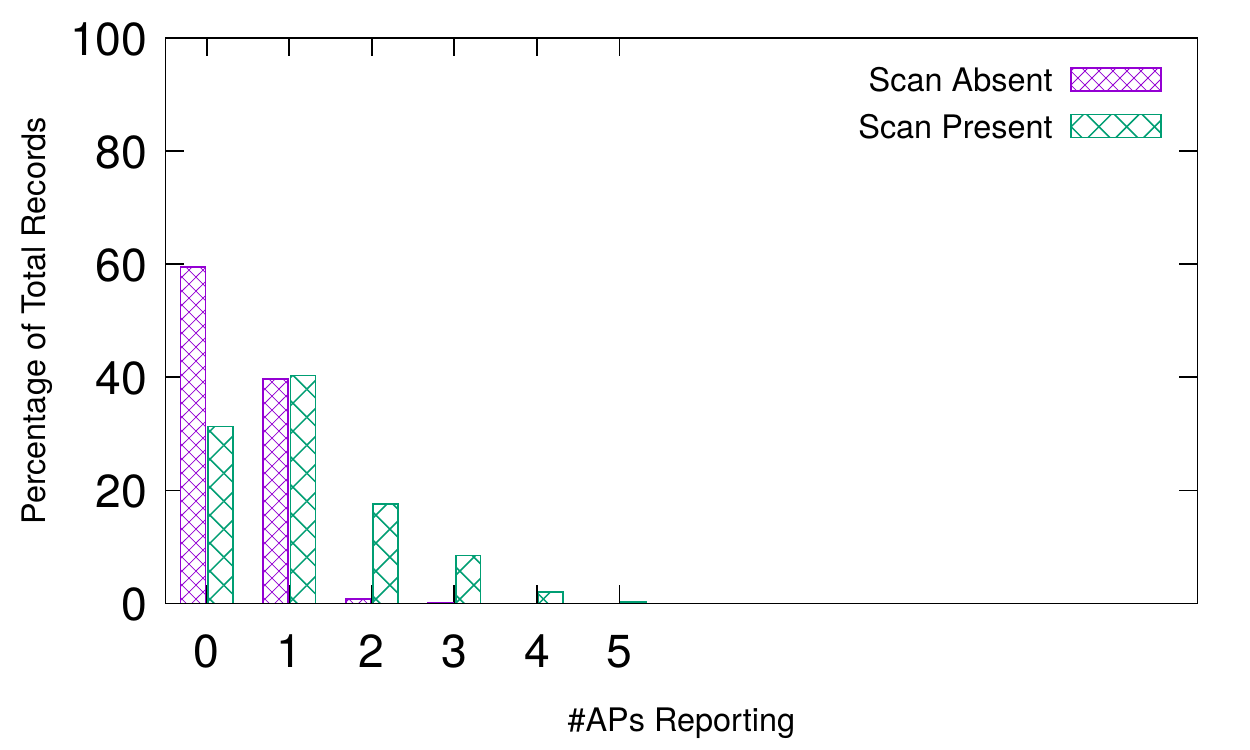}
\\ (g) \textbf{State: Active - $2.4$ GHz} & (h) \textbf{State: Active - $5$ GHz}\\
\end{tabular}

\caption{The number of APs reporting in the absence and presence of scanning for
$4$ client states -- Disconnected, Inactive, Intermittent, and Active. When a client is disconnected, the APs report only when the client scans (Figure ($a$) and ($b$)). For the remaining $3$ states, the number of reporting APs are present irrespective of the scanning status; however, the amount of APs are consistently higher when the client scans. $2.4$ GHz always has more APs reporting than $5$ GHz. }
\label{Fig:cardinality_scanvsnoscan_statewise}

\end{figure*}

In the absence of client scans, APs get only non-scanning frames.
For each of the four states of the client, we study how many APs report that client \textit{i.e.} the cardinality. 
With this analysis, we are able to compare the cardinality in the presence and absence of scans, for both $2.4$ and $5$ GHz bands.
We see that as the frequency of scanning increases, more number of APs respond and the cardinality increases.
We expand the result for each state of the client in Figure~\ref{Fig:cardinality_scanvsnoscan_statewise}.
The cardinality is consistently higher for $2.4$ GHz than that of $5$ GHz, that further increases in the presence of scans.
This implies that higher frequency of scanning possibly reduces the Cardinality Mismatch and vice-versa.

\begin{figure}[t]
 \centering
 \includegraphics[scale=0.6]{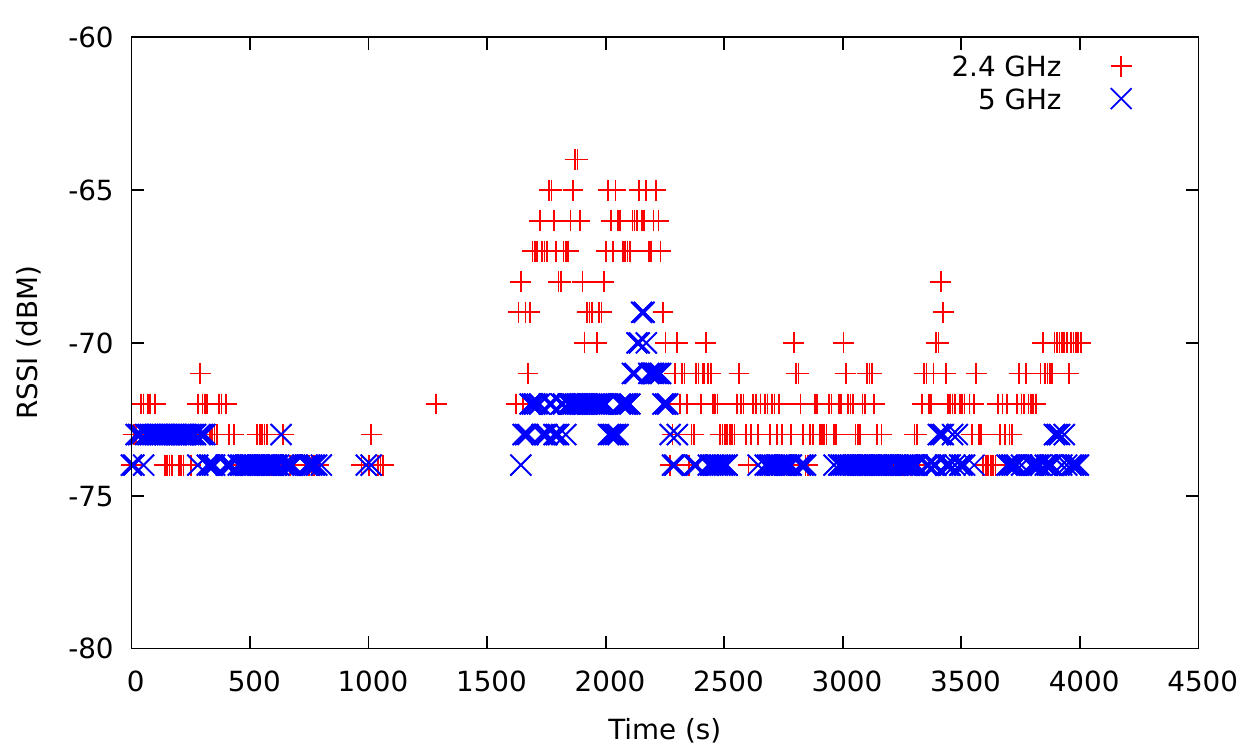}
 \caption{Variations in RSSI in $2.4$ GHz and $5$ GHz for a stationary client as measured at the server. Notice that $5$ GHz is relatively stable than $2.4$ GHz.}
\label{Fig:RSSIFluctuations-24vs5}
\end{figure}%

\begin{table}[t]
\begin{subtable}{.5\textwidth}
\scriptsize
\centering
\begin{tabular}{@{} m{1.5cm}m{2cm}m{1.5cm}m{2cm} @{}}
\toprule
\textbf{Frames} & \textbf{Transmission Distance} & \textbf{RSSI Variation} & \textbf{Frequency of Transmission}\\
\hline
Scanning  & \emph{High} -- \cmark & \emph{Low} -- \cmark & \emph{Low} -- \xmark \\
Non-Scanning & \emph{Low} -- \xmark & \emph{High} -- \xmark & \emph{High} -- \cmark \\
\bottomrule
\end{tabular}

\end{subtable}
\\
\begin{subtable}{.5\textwidth}
\scriptsize
\centering
\begin{tabular}{@{} m{1.5cm}m{2cm}m{1.5cm}m{2cm}  @{}}
\toprule
\textbf{Band} & \textbf{Transmission Distance} & \textbf{RSSI Variation} & \textbf{Frequency of Scanning}\\
\hline
$2.4$ GHz & \emph{High} -- \cmark & \emph{High} -- \xmark & \emph{High} -- \cmark \\
$5$ GHz & \emph{Low} -- \xmark & \emph{Low} -- \cmark & \emph{Low} -- \xmark \\
\bottomrule
\end{tabular}
\end{subtable}
\caption{A summary of the causes and their impact (\cmark - Reduces Localization Errors, \xmark - Increases Localization Errors). The causes conflict with each other, making server-side localization non-trivial.}

\label{tbl:observations_summary}
\end{table}

However, a downside for $2.4$ GHz is that the frames (both scanning and non-scanning) 
have higher variation in RSSI.
This means, even though the extent of the Cardinality Mismatch is lower, 
the RSSI will differ more in $2.4$ GHz.
To confirm this, we analyze the RSSI from a stationary client by enabling its association in one band at a time and disabling the other band altogether. We use the RSSI recorded at the RTLS server for a duration of $1$ hour.
As shown in Figure~\ref{Fig:RSSIFluctuations-24vs5}, even with more scanning information available $2.4$ GHz is more prone to RSSI fluctuations than $5$ GHz. 
The reasons for this behavior are ($a$) the range of $2.4$ GHz is almost double than that of $5$ GHz 
and ($b$) a lesser number of non-overlapping channels makes it susceptible to interference. 
Therefore, RSSI from $2.4$ GHz results in predicting distant and transient locations.
We validate this in different locations with devices of four other models.

To summarize, there is a significant extent of Cardinality Mismatch and High Client Scan Latency.
There is a difference in the extent of the issues for the two classes of frames 
and the two bands of operation.
While scanning frames has a longer distance of transmission and 
less variation in RSSI, they are not often sent by the clients.
The factors favoring $2.4$ GHz are longer distance of transmission and
higher frequency of scanning.
However, low variation in RSSI works in favor of $5$ GHz.
We summarize these observations in Table~\ref{tbl:observations_summary}.

\begin{figure*}[t]
\centering
        \begin{subfigure}[b]{0.46\textwidth}
        \centering
		\includegraphics[scale=0.3]{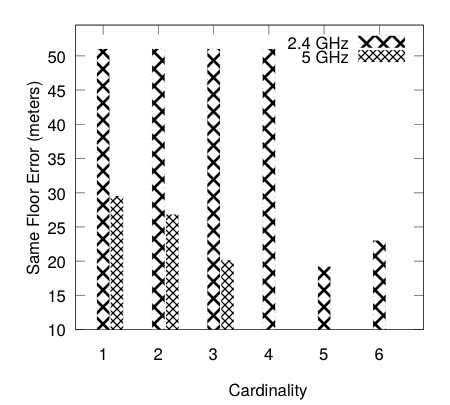}
                \caption{Same Floor}
                \label{Fig:Baseline-SameFloor}
        \end{subfigure}%
        \qquad
        \begin{subfigure}[b]{0.46\textwidth}
        \centering
		\includegraphics[scale=0.3]{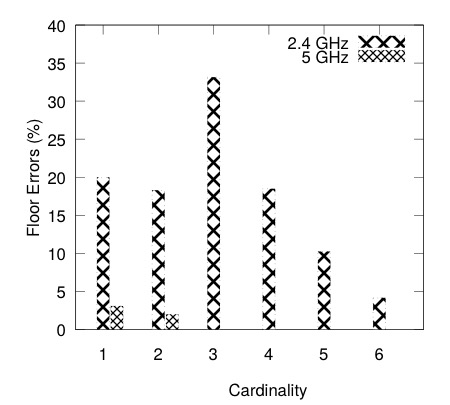}
                \caption{Different Floor}
                \label{Fig:Baseline-FloorErrors}
        \end{subfigure}%
\caption{Localization errors reported with baseline fingerprinting algorithm. The errors are measured as Same Floor errors in meters and Different Floor errors in percentage of floors for Cardinalities ranging from $1$ to $6$. 
Cardinalities \textgreater $3$ are Not Applicable to $5$
GHz due to cardinality mismatch.}\label{Fig:accuracy_results_baseline}

\end{figure*}

\subsection{Impact of Causes on Localization Errors}
We now evaluate the impact of the causes on the localization errors.
We implemented a server-side localization using well known fingerprint-based method~\cite{radar}. 
Since we use server-side processing, we do not require any client-side modification.
Our proposals do not make assumptions about hardware or operating system of the clients or the controller.
Although each adaptation of fingerprint-based technique from the existing body of work may result in different errors,
our exercise gives us a baseline that cuts across all the adaptations. 
The $2.4$ and $5$ GHz bands differ in distance of transmission, variation in RSSI, and frequency of scanning.
We measure the localization errors for both the bands.

We report localization errors for each value of the cardinality in online phase to 
understand how the error varies as a function of the cardinality.
We consider a multi-storey building; hence, the clients may be localized across floors, irrespective of their current position.
Therefore, we measure the errors in terms of ($a$) Different Floor and ($b$) Same Floor errors.
Different Floor error is the percentage of total records for which a wrong floor is estimated.
These are seen at the higher percentiles.
For the rest of the records, the Same Floor error is the distance in meters 
between the actual and the predicted landmark on the floor.
The errors at the higher percentiles are essential for security applications, for example, an error by a floor in localizing a suspect can make or break the evidence. 
We want to minimize both the errors.

Figure~\ref{Fig:accuracy_results_baseline} shows the results for the Baseline fingerprinting-based localization. Overall, we observe that the errors are high for the low cardinalities, 
and the errors for $2.4$ GHz are more significant compared to that of $5$ GHz.
This is despite the fact that more APs hear clients and
frequency of transmission of scanning frames is more for $2.4$ GHz. We conclude that the variation in RSSI, including that induced by the transmission power control,
has a significant impact on the cardinality and therefore on the localization errors.

\begin{figure*}[t]
\centering
        \begin{subfigure}[t]{\linewidth}
        \centering
		\includegraphics[scale=0.3]{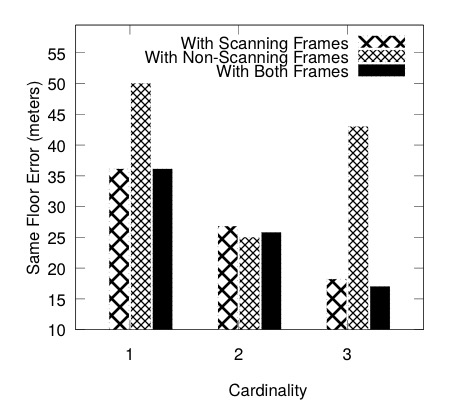}
		\,
		\includegraphics[scale=0.3]{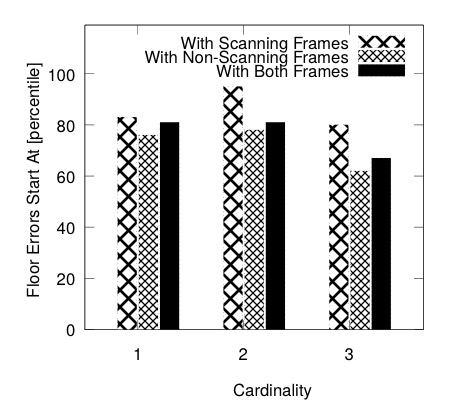}
        \caption{2.4 GHz}
        \label{Fig:LocalizationErrors75th-2.4GHz}
        \end{subfigure}%
        
        \begin{subfigure}[t]{\linewidth}
        \centering
		\includegraphics[scale=0.3]{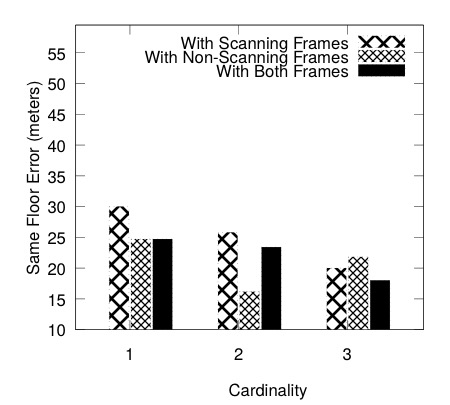}
		\,
        \includegraphics[scale=0.3]{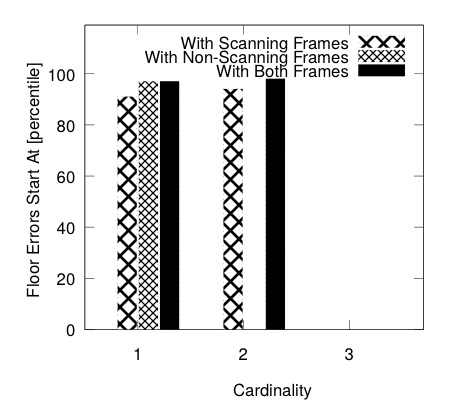}
        \caption{5 GHz}
        \label{Fig:LocalizationErrors75th-5GHz}
        \end{subfigure}%
\caption{Bifurcation of Localization Errors with and without scanning frames for 2.4 and 5 GHz. Localizing client with only scanning frames reduces errors in 2.4 GHz; while only non-scanning frames reduces errors in 5 GHz.}
\label{Fig:localizationerrors_scanNoScan}

\end{figure*}

In order to understand the significance of scanning frames on localization errors, we study the spectrum of errors caused due to each type of frame -- scanning and non-scanning. Results in Figure~\ref{Fig:accuracy_results_baseline}  combine both types of frames; we now bifurcate the results for each frame type. 
The analysis is categorized for  -- ($a$) both scanning and non-scanning frames, ($b$) only scanning frames, and ($c$) only non-scanning frames. 
Figure~\ref{Fig:localizationerrors_scanNoScan} shows the localization errors on the same floor and different floors for both frequency bands. 

Scanning frames not only reduce the localization errors on the same floor in $2.4$ GHz but, even floor errors start at as high as 85$^{th}$\%ile. With non-scanning frames, floor errors start at as low as 62$^{th}$\%ile. The reason being that the scanning frames, as opposed to non-scanning frames, do not incorporate transmit power control, implying lesser or no RSSI variation, which in turn helps in a close match of online and offline fingerprints. This improves accuracy in $2.4$ GHz. While non-scanning frames increase the localization errors in $2.4$ GHz. The reason being that the non-scanning frames fetch RSSI from AP to which the client is associated and APs which are on the same or overlapping channels as the AP of association. A client connects to a farther AP because ($a$) in $2.4$ GHz, a farther AP can temporarily have better RSSI, ($b$) algorithms, such as load balancing, cause overloaded APs, which may be nearer to the client, to not respond while farther APs respond, and ($c$) sticky clients, these clients do not disassociate with an AP of low RSSI, which is a limitation of client drivers.

The results for the $5$ GHz are contrasting to that of $2.4$ GHz.
Non-scanning frames reduce the localization errors in $5$ GHz.
Important to note here is that the 5 GHz band hardly encounters floor errors, as is seen in Figure~\ref{Fig:LocalizationErrors75th-5GHz}.
Primary reason being the $2$x reduced range of $5$ GHz ($15$ meters) when compared to $2.4$ GHz($30$ meters). 
Non-scanning frames in $5$ GHz fetch RSSI from the AP to which the client is associated, which is the strongest AP. 
Hence, the RSSI from this AP have minimal variation due to reduced interference. Therefore, they match closely with fingerprint maps and improve reduce errors.
However, the scanning frames in $5$ GHz do not fetch RSSI from the AP to which the client is associated, that is the strongest AP. RSSI from other APs, with which the client is not associated, may or may not closely match with offline fingerprints. Hence, the accuracy suffers.

We analyze the localization errors when a mix of scanning and non-scanning frames are used for localizing the client.
In this case, the errors range between the numbers reported by only scanning and only non-scanning frames. 
Overall, we see that the $5$ GHz outperforms $2.4$ GHz, irrespective of which type of frames are considered for the localization.
However, accuracy of localization in $2.4$ GHz is severely affected by the minimal scans.
Thus, preferring scanning frames in $2.4$ GHz and non-scanning frames in $5$ GHz is beneficial to improve localization accuracy.

\begin{figure*}[t]
\centering
        \begin{subfigure}[t]{\linewidth}
        \centering
		\includegraphics[scale=0.3]{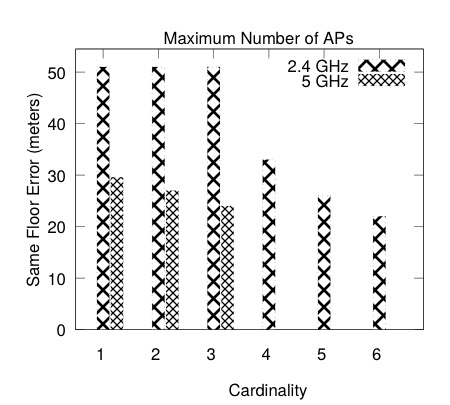}
		\,
		\includegraphics[scale=0.3]{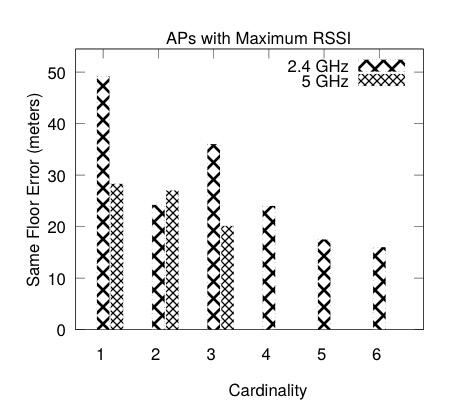}
		\,
		\includegraphics[scale=0.3]{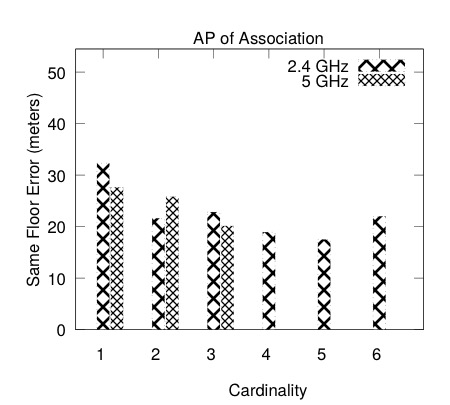}
        \caption{Same Floor Errors}
        \label{Fig:heuristics-samefloor}
        \end{subfigure}%
        \\
        \begin{subfigure}[b]{\linewidth}
        \centering
		\includegraphics[scale=0.3]{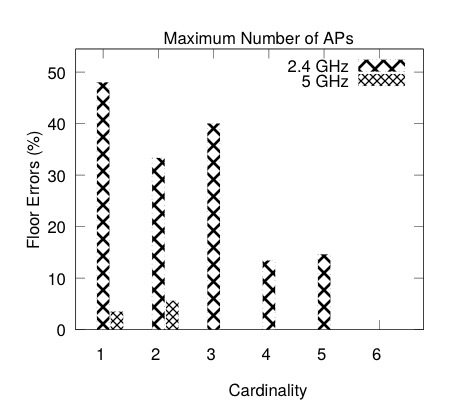}
		\,
        \includegraphics[scale=0.3]{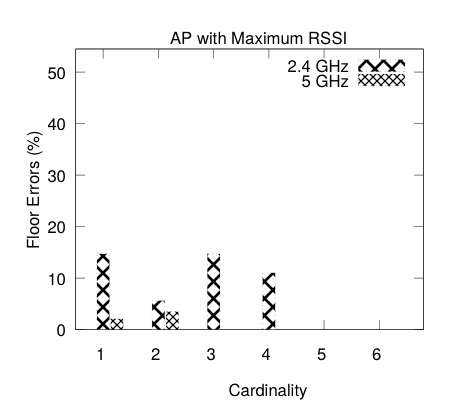}
		\,
        \includegraphics[scale=0.3]{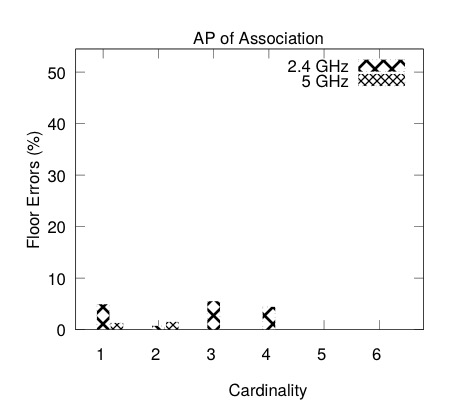}
        \caption{Different Floor Errors}
        \label{Fig:heuristics-differentfloor}
        \end{subfigure}%
\caption[Localization Errors with Three floor detection heuristics.]{The localization error with three floor detection heuristics. The lowest localization errors were seen for the heuristic -- AP of Association. On an average, across all the cardinalities, AP of association heuristic reduces the same floor and different floor errors by $58$\% and $78$\% in $2.4$ GHz; $3.8$\% and $46$\% in $5$ GHz, respectively. Note that cardinalities greater than 3 are not applicable in 5 GHz.}
\label{Fig:localizationerrors_heuristics}

\end{figure*}

\subsection{Reducing the Localization Errors}
\label{sec:solution}

The challenge in large-scale deployments is that clients or APs cannot be modified. So, the odds of installing mobile apps that can trigger scans are very low. Furthermore, the latest phones do not prefer, for the purposes of saving bandwidth, to trigger scans until absolutely necessary. Therefore, even if the scanning frames have a potential of improving the localization accuracy in 2.4 GHz, their frequency is not in one's control.
Since all the identified causes in Table~\ref{tbl:observations_summary}, are conflicting to each other, getting rid of the two %
issues -- Cardinality Mismatch and Low Cardinality, is not trivial.

For improvement, we take a position to make the best use of whatever RSSI we receive during the online phase.
We use heuristic to select the APs from the online phase to reduce localization errors.
We know the location of each AP. 
We use this information to shortlist the APs from the online fingerprint. 
The algorithm first selects a floor and then shortlists all the APs 
that are located on the same floor.
We use the shortlisted APs to find a match with offline fingerprints. 
For selecting the floor, we explore three heuristics -- 
($a$) Maximum Number of APs - the floor for which the maximum APs are reporting the client, 
($b$) AP with Maximum RSSI - the floor from which the strongest RSSI is received, and 
($c$) AP of Association - the floor of AP to which the client is presently associated with.

\begin{figure}[t]
\centering
\includegraphics[scale=0.6]{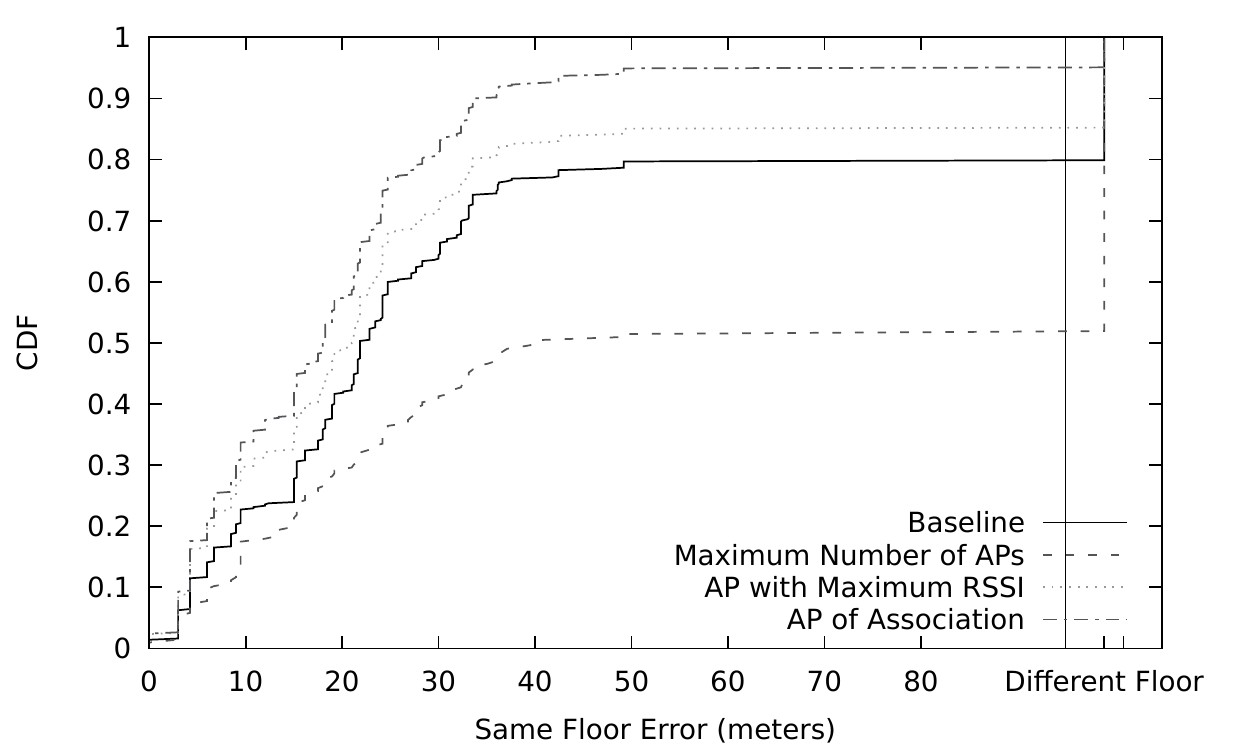}
     
\caption{Localization errors with three 
floor detection heuristics -- ($a$) Maximum Number of APs, ($b$) AP with Maximum RSSI, and ($c$) AP of Association at $Cardinality$=$1$ for $2.4$ GHz. AP of Association performs the best for both bands. Maximum Number of APs performs worse than the Baseline, while the other two significantly reduced the errors.}
\label{Fig:Card1-Accuracy-CDF}
\end{figure}

Figure~\ref{Fig:localizationerrors_heuristics} shows how the localization errors vary for the three heuristics for $85^{th}$ percentile values.
There is clearly an improvement for both Same Floor and Different Floor errors.
Floor detection with Maximum Number of APs gives the least improvement. In fact, until cardinality $4$ it performs worse than the Baseline.
A cause behind this is that the distant APs, specially in $2.4$ GHz that has longer transmission distance, 
respond and thus localization errors increase.
Next,
is the floor detection with the AP with Maximum RSSI and AP of Association.
The AP with Maximum RSSI or the AP of Association are typically closest to the client, 
except when the controller does load balancing and transmit power control.
There is marginal improvement for $5$ GHz.
Since the Figure~\ref{Fig:localizationerrors_heuristics} only shows the data for $85^{th}$ percentile, 
we plot the CDF of error for cardinality=$1$ in Figure~\ref{Fig:Card1-Accuracy-CDF} for $2.4$ GHz.
We see that the error reduces for all the percentiles. 
We see similar results for other cardinalities, but we don't include them due to space constraint.

We compare our results with Signal SLAM~\cite{signalSLAM} which is deployed in a public space like mall since we also have similar deployments. We have similar observations in other venues too. 
We find their $90^{th}$ percentile is about $15$ meters.  We perform similar. 
In fact, their AP visibility algorithm has $90^{th}$ percentile as $24.3$
meters.  We perform better than this in $5$ GHz. 
 Given the complexity of the algorithm Signal SLAM incorporates and the amount of sensing it needs, we believe
even with few meters of accuracy our approach is better; particularly
because its simple and scalable.

\section{Discussion}
\label{sec:discussion}
Now, we discuss the practical challenges encountered while localizing clients in real deployments and limitations of our solution.

\subsection{Challenges Of Real Deployments}
Real deployments have myriad of practical challenges that hamper the efficiency and the accuracy of an empirical study.
For instance, there can be sudden and unexpected crowd movement which is known to increase signal variations.
Furthermore, as and when required network administrators either replace old APs or deploy new APs. These administrative decisions are not under our control. However, such changes severely affect the offline fingerprints and change the floor heat maps that ultimately affect location accuracy.
Preparing fingerprints for an entire campus with several thousands of landmarks is already tedious, such developments make the process of iterations even more cumbersome.

Beyond insufficient measurement and latency issues, various contextual dynamics makes the fingerprint-based system erroneous. The primary reason is that such dynamic changes results in significant fluctuation in RSSI measurements, which affects the distance calculation of the localization algorithms.
These fluctuations can happen quite frequently as there are many different factors affecting RSSI between an AP and its clients, such as crowds blocking the signal path, AP-side power-control for load balancing, and client-side power control to save battery. In Section~\ref{sec:challenges} we shed light on most of these factors.
Lastly, all MAC addresses in our system are anonymized. We do not do a device to person mapping to preserve user privacy.

\subsection{Limitations}
A major limitation of this work is that we have not considered an exhaustive set of devices. 
Given a multitude of device vendors, it is practically impossible to consider all set of devices for this kind of in-depth analysis. We did cover the latest set of devices, though, including iPhone and Android devices.
The second limitation is that even though we collected the data for both lightly (semester off, very few students on campus) and heavily loaded (semester on, most students on campus) days. We tested our observations on the lightly loaded dataset but, only on a subset of heavily loaded days. We do not yet know the behavior of system during heavily loaded days, in its entirety. Specifically, the load, concerning the number of clients and traffic is expected to increase interference and thus, signal variations. However, this study is still a part of future work.
The third limitation of this work is that we do not consider the effect of MAC address randomization algorithms which make clients intractable. Although there is an active field of research that suggest ways to map randomized MAC to actual MAC~\cite{MACRandomization}, but given its complexity we do not employ these.

\section{Conclusion}\label{sec:conclusion}

To conclude, we presented two major issues that need to be addressed to perform server-side localization.
We validated these challenges with a huge data from a production WLAN deployed across a university campus.
We discussed the causes and their impact on these challenges.
We proposed heuristics that handle the challenges and reduce the localization errors.
Our findings apply to all the server-side localization algorithms, which use fingerprinting techniques.
Most of this work provides real-world evidence of ``where'' and ``what'' may go wrong for practically localizing clients in a device agnostic manner.

\bibliographystyle{ACM-Reference-Format}
\bibliography{references.bib}

\end{document}